\documentclass[11pt,a4paper]{article}
\pdfoutput=1

\usepackage{jcappub}
\makeatletter
\gdef\@fpheader{}
\g@addto@macro\bfseries{\boldmath}
\makeatother

\usepackage[utf8]{inputenc}
\usepackage{graphicx}
\usepackage{epsfig}
\usepackage{subfigure}
\usepackage{color}
\usepackage{comment}

\newcommand{\ie}{{i.e.~}}

\newcommand{\eg}{\textsl{e.g.~}}

\newcommand{\order}[1]{\mathcal{O}\!\left(#1\right)}

\newcommand{\dd}{\mathrm{d}}
\newcommand{\ee}{e}

\newcommand{\sss}[1]{{\scriptscriptstyle{#1}}}

\newcommand{\uPl}{\mathrm{Pl}}

\newcommand{\uend}{\mathrm{end}}

\newcommand{\ueff}{\mathrm{eff}}

\newcommand{\usssPl}{\sss{\uPl}}

\newcommand{\calP}{\mathcal{P}}

\newcommand{\GeV}{\mathrm{GeV}}

\newcommand{\Mp}{M_\usssPl}

\newcommand{\efolds}{$e$-folds~}

\newcommand{\beq}{\begin{equation}}
\newcommand{\eeq}{\end{equation}}
\newcommand{\bea}{\begin{eqnarray}}
\newcommand{\eea}{\end{eqnarray}}

\newlength{\wsingfig}
\newlength{\wdblefig}
\newlength{\wquadfig}
\newlength{\wtriplefig}
\newcommand{\Eq}[1]{Eq.~(\ref{#1})}
\newcommand{\Eqs}[1]{Eqs.~(\ref{#1})}
\newcommand{\Fig}[1]{Fig.~{\ref{#1}}}
\newcommand{\Figs}[1]{Figs.~{\ref{#1}}}
\newcommand{\Ref}[1]{Ref.~{\cite{#1}}}
\newcommand{\Refs}[1]{Refs.~{\cite{#1}}}
\newcommand{\Sec}[1]{Sec.~\ref{#1}}
\newcommand{\Secs}[1]{Secs.~\ref{#1}}
\newcommand{\App}[1]{Appendix~\ref{#1}}

\title{A novel way to determine \\ the scale of inflation}

\author[a]{Kari Enqvist,}
\author[b]{Robert J. Hardwick,}
\author[c]{Tommi Tenkanen,}
\author[b,d]{\\ Vincent Vennin}
\author[b]{and David Wands}

\affiliation[a]{Department of Physics, University of Helsinki and Helsinki Institute of Physics, \\
                      P.O.~Box 64, FI-00014, Helsinki, Finland}
\affiliation[b]{Institute of Cosmology \& Gravitation, University of Portsmouth, Dennis Sciama Building, Burnaby Road, Portsmouth, PO1 3FX, United Kingdom}
\affiliation[c]{Astronomy Unit, Queen Mary University of London, \\
                      Mile End Road, London, E1 4NS, United Kingdom}
\affiliation[d]{Laboratoire Astroparticule et Cosmologie, Universit\'{e} Denis Diderot Paris 7, \\
                      10, rue Alice Domon et L\'eonie Duquet, 75013 Paris, France}

\emailAdd{kari.enqvist@helsinki.fi}
\emailAdd{robert.hardwick@port.ac.uk}
\emailAdd{t.tenkanen@qmul.ac.uk}
\emailAdd{vincent.vennin@apc.in2p3.fr}
\emailAdd{david.wands@port.ac.uk}

\abstract{
We show that in the Feebly Interacting Massive Particle (FIMP) model of Dark Matter (DM), one may express the inflationary energy scale $H_*$ as a function of three otherwise unrelated quantities, the DM isocurvature perturbation amplitude, its mass and its self-coupling constant, independently of the tensor-to-scalar ratio. The FIMP model assumes that there exists a real scalar particle that alone constitutes the DM content of the Universe and couples to the Standard Model via a Higgs portal. We consider carefully the various astrophysical, cosmological and model constraints, accounting also for variations in inflationary dynamics and the reheating history, to derive a robust estimate for $H_*$ that is confined to a relatively narrow range. We point out that, within the context of the FIMP DM model, one may thus determine $H_*$ reliably even in the absence of observable tensor perturbations.
}

\keywords{}

\begin{document}
\maketitle
\section{Introduction}
The history of the Universe is well understood up to energies $T \sim$ MeV, where Big Bang Nucleosynthesis (BBN) occurred. At this point the Universe was: spatially flat; homogeneous; isotropic to a high degree; and its baryonic matter content was in thermal equilibrium with different patches exhibiting only very small deviations from the average temperature, $\delta T/T \sim 10^{-5}$. 

The current paradigm for explaining, not only why the early Universe had these properties, but also the origin of the small inhomogeneities --- that later became the seeds for the large scale structure of the Universe --- is cosmic inflation~\cite{Starobinsky:1980te, Sato:1980yn, Guth:1980zm, Linde:1981mu, Albrecht:1982wi, Linde:1983gd}: an era of accelerated expansion before the standard Hot Big Bang state of the Universe. Cosmic inflation is a highly successful explanation for these aspects, and many of its phenomenological realisations have been studied in the literature over the last few decades (for major reviews of inflation models, see \Refs{Lyth:1998xn,Mazumdar:2010sa,Martin:2013tda,Patrignani:2016xqp}).

Amongst the parameters that are relevant to inflationary perturbations, two have been measured: the amplitude of the curvature power spectrum, $A_s$, and the corresponding spectral tilt, $n_s$, which the {\it Planck} collaboration have recently measured to an accuracy of $\Delta A_s = \mathcal{O}(10^{-2})$ and $\Delta n_s = \mathcal{O}(10^{-3})$ \cite{Ade:2015xua}. However, the energy scale at which inflation --- or more accurately, the last $\sim 60$ \efolds of inflation --- happened is still unknown. The energy scale of inflation can be characterised by the value of the Hubble parameter during inflation, $H_*$. In single-field slow-roll models of inflation, this can be expressed by the primordial tensor-to-scalar ratio, $r$, as $H_* = 8\times 10^{13}\sqrt{r/0.1}$ GeV. The current upper bound provided by the joint analysis of {\it BICEP2/Keck Array} and {\it Planck} data is $r<0.12$ \cite{Ade:2015tva}, whereas no strict lower bound exists other than the requirement for realising successful BBN at $T\sim 1$ MeV \cite{Kawasaki:2000en,Hannestad:2004px,Ichikawa:2005vw,DeBernardis:2008zz}. Hence, there is a huge gap between the scales at which the dynamics of the Universe is understood. It is elementary then, and of great importance to understanding the physics between these scales, to quantify how large the gap is.

The next-generation experiments may be able to push the upper bound for the tensor-to-scalar ratio down to $r<0.03$ from {\it BICEP3} \cite{Wu:2016hul} and $r<0.001$ from {\it LiteBIRD} \cite{Matsumura:2013aja} or {\it COrE} \cite{Finelli:2016cyd, DiValentino:2016foa}, or any of these may detect it above these limits. However, these numbers illustrate that if no detection is made, even in the best possible case the planned experiments cannot determine the inflationary scale by primordial tensor modes if it was smaller than $H_*\simeq 8\times 10^{12}$ GeV. It would therefore be interesting if one could find scenarios in which the inflationary scale could be determined by other means. This is our aim in the present paper.

Based on the work conducted by one of the present authors and collaborators in \Refs{Nurmi:2015ema,Kainulainen:2016vzv,Heikinheimo:2016yds}, we present a scenario where the scale of inflation $H_*$ is determined by three observables: the dark matter (DM) isocurvature perturbation amplitude, its mass and self-coupling constant. This determination is made completely independently of the tensor-to-scalar ratio $r$, increasing the range in $H_*$ that one can infer to values for the inflationary scale well below the current lower bound, or below the sensitivity of the next-generation experiments. Furthermore, we find that in this scenario the inflationary scale can be determined almost solely from {\it spectator field} dynamics, i.e. from the dynamics of a scalar field which was a light (effectively massless) test (energetically sub-dominant) field during inflation, and weakly depends on the dynamics of the field(s) driving inflation itself. 

As a representative example of this kind of scenario, we study a generic real singlet scalar extension to the Standard Model of particle physics (SM). The new singlet scalar particle is a Feebly-Interacting Massive Particle (FIMP) \cite{McDonald:2001vt,Hall:2009bx,Bernal:2017kxu}, which we assume to constitute the DM abundance\footnote{For other scenarios connecting FIMP DM and inflation, see Refs. \cite{Bernal:2017kxu,Tenkanen:2016twd,Heurtier:2017nwl,Argurio:2017joe}.}. Due to a feeble coupling between the singlet scalar and the SM sector, the singlet never thermalises with the SM and the DM abundance is produced by the ``freeze-in'' mechanism instead of the standard freeze-out. We discuss this in detail throughout the following sections.

The paper is organised as follows. In \Sec{sec:basic-argument} we provide a general outline of the scenario used to make the inference of the inflationary scale $H_*$. We start by considering a minimal scenario and derive a basic formula for $H_*$ using the essential features of the model that combine the constraints on the primordial isocurvature perturbations and the DM self-interaction cross-section. The main focus of the following \Sec{extended_scenario} is to identify as many potential degeneracies within the $H_*$ formula as possible due to variations within the inflationary dynamics (\Sec{sec:vary-inflation}), reheating history (\Sec{sec:vary-reheating}), and low-energy particle dynamics (\Sec{sec:low-energy-dynamics}). Finally, in \Sec{sec:conclusions}, we provide a concluding discussion.

\section{Minimal scenario: the basic argument} \label{sec:basic-argument}

We begin by presenting a simple version of the scenario where the energy scale of inflation can be determined without measuring the tensor-to-scalar ratio. The model we consider is a minimal extension to the SM Lagrangian, where in addition to the SM particle content there is a $\mathbb{Z}_2$-symmetric real singlet scalar, $s$, coupled to the SM via the Higgs portal \cite{Silveira:1985rk,McDonald:1993ex}
\begin{equation} \label{eq:model-potential}
\ \mathcal{L} = \mathcal{L}_{\rm SM} 
-\frac12 \partial^\mu s \partial_\mu s + \frac{m_s^2}{2}s^2 + \frac{\lambda_s}{4}s^4 + \frac{\lambda_{hs}}{2}\Phi^\dagger \Phi s^2\, . 
\end{equation}
In this expression, $\mathcal{L}_{\rm SM}$ is the SM Lagrangian\footnote{Radiative corrections in a curved background generate an extra term to the scalar potential, $\ V_{\rm G} = \xi_h h^2 R  + \xi_s s^2 R$, constituting of the non-minimal couplings to gravity $\xi_h$, $\xi_s$ of both the Higgs and singlet, respectively \cite{Callan:1970ze,Freedman:1974gs}. For this scenario, we shall consider the case where the singlet has negligible $\xi_s$. The value of the SM Higgs non-minimal coupling to gravity is not relevant for our purposes. We also assume that the potential issue concerning the instability of the Higgs potential occuring roughly at the scale $10^{11}$ GeV within the SM \cite{Buttazzo:2013uya} is fixed by some new physics which decouples from the $s$ sector, for example a coupling between the Higgs field and the inflaton \cite{Lebedev:2012sy}, which is in any case necessary to reheat the Universe after inflation.} and the SM Higgs doublet in the unitary gauge is written as $\sqrt{2}\Phi^{\rm T} = (0,v + h)$, where $v$ is the vacuum expectation value of the Higgs field. We assume that the portal coupling takes a small value\footnote{Note that this does not impose a fine-tuning issue, as the running of the portal coupling is always very small in this model \cite{Alanne:2014bra,Heikinheimo:2017ofk}.}, $\lambda_{hs}<10^{-7}$, so that the singlet $s$ does not thermalise with the SM in the early Universe, but remains a FIMP DM candidate \cite{McDonald:2001vt, Hall:2009bx}. 
The $s$ particles can constitute all the DM if the Higgs field can produce sufficient number of $s$ particles from Higgs decay after electroweak symmetry breaking or, if the decay is not kinematically allowed, if Higgs-mediated gauge boson annihilations into $s$ particles are frequent enough \cite{McDonald:2001vt, Yaguna:2011qn,Bernal:2017kxu}. 
For the basic scenario, the exact production mechanism is not relevant, and we will discuss the low-energy dynamics in more detail in \Sec{sec:low-energy-dynamics}.

Let us see how to determine the scale of inflation with the known behaviour of spectator fields during inflation. During inflation, light fields ($\partial^2V/\partial s^2 \ll H^2$, where $V$ is the scalar potential and $H$ is the Hubble scale) generally develop long (super-Hubble) wavelength excitations leading to a non-zero variance. The field approaches a stationary distribution \cite{Starobinsky:1986fx} characterised by $\langle V \rangle \sim H^4$ for a sufficiently slowly-varying Hubble rate~\cite{Enqvist:2012xn, Enqvist:2014zqa, Hardwick:2017fjo}, with a typical value
\begin{equation} 
\label{eq:stationary_s}
\left. s \right\vert_{\rm typical} \simeq \sqrt{\langle s^2\rangle} =  \left[ \frac{3}{2\pi^2\lambda_s}\dfrac{\Gamma^2\left(\frac{3}{4}\right)}{\Gamma^2\left(\frac{1}{4}\right)} \right]^{\frac{1}{4}} H\, ,
\end{equation}
where $\langle \cdot \rangle$ denotes the ensemble-average over separate Hubble-sized patches. In deriving \Eq{eq:stationary_s} we require that the quartic terms in the scalar potential dominate over the quadratic ones, $\lambda_s \langle s^2\rangle \gg 2m_s^2+\lambda_{hs}\langle h^2\rangle$; we will verify that this is always the case in \Sec{sec:vary-inflation}.

\begin{figure}[h!]
\begin{center}
\includegraphics[width=0.8\textwidth]{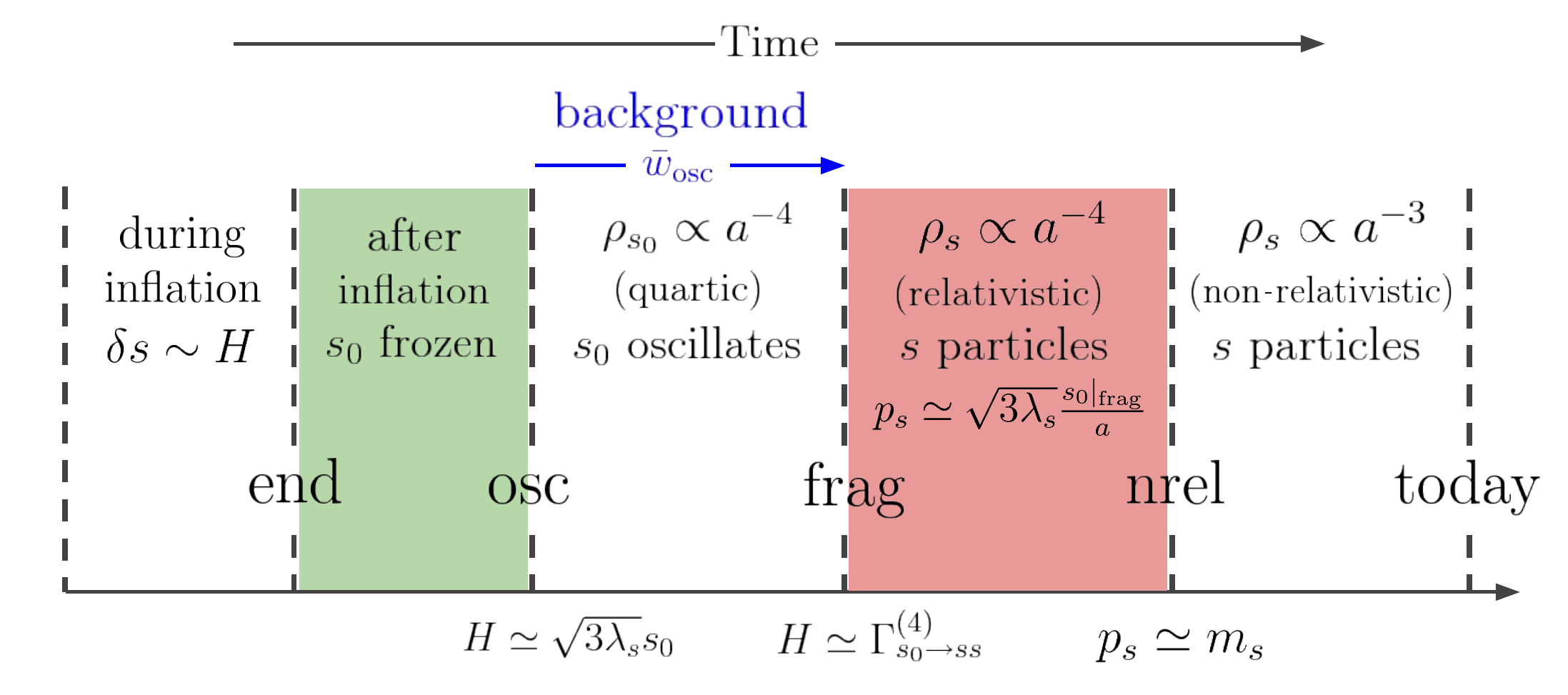}
\caption{\label{fig:reheating} Timeline for the dynamics of the singlet scalar field studied in this paper. During inflation, $\delta s\sim H$ refers to the typical size of fluctuations during inflation. After inflation, when the Hubble rate $H$ drops below the effective mass of the scalar field $\sqrt{3\lambda_s}s_0$, it starts to oscillate at the bottom of its quartic potential, and $\rho_{s_0}$, the energy density of the singlet condensate, decays as $1/a^4$. The mean equation-of-state parameter of the background energy density during this oscillation period is denoted $\bar{w}_{\rm osc}$. When $H$ drops below the fragmentation rate $\Gamma_{s_0\rightarrow ss}^{(4)}$, the singlet condensate fragments into singlet $s$ particles with typical momentum $p\simeq \sqrt{3\lambda_s} s_0|_{\rm frag}$ that redshifts as $1/a$. When this momentum reaches the mass $m_s$, these particles become non-relativistic, and $\rho_s$, the energy density contained in the singlet particles, decays as $1/a^3$. Its final value in this multi-stage process, which determines the DM abundance, is derived in \App{app:general-scaling-energy-density}.}
\end{center}
\end{figure}

Both the Higgs and $s$ field fluctuations represent isocurvature perturbations relative to the adiabatic inflaton perturbations during inflation\footnote{Unless one of them is the inflaton, but in this paper we do not consider this possibility.}. Soon after inflation the Universe becomes radiation-dominated; once the Hubble rate drops below their effective mass the fields start to oscillate about their minima. The Higgs field then decays into radiation quickly, typically within a few e-folds~\cite{Figueroa:2015rqa}, reaching thermal equilibrium and thus leaving only adiabatic perturbations in the SM radiation. However, due to the feeble coupling between the singlet scalar and the SM, the $s$ condensate (denoted by $s_0$ from now on) does not thermalise and therefore its fluctuations remain isocurvature perturbations relative to the adiabatic perturbations of the SM radiation. Even though the $s_0$ condensate is assumed not to decay into SM radiation, the condensate may fragment into $s$ particles which eventually become cold (non-relativistic) DM particles and inherit the primordial isocurvature perturbations from the condensate. This happens if $\lambda_s$ is large enough, so that the $s_0$ condensate fragments while still in an effectively quartic potential \cite{Nurmi:2015ema,Kainulainen:2016vzv}, and this condition will be carefully checked in \Sec{sec:vary-reheating}. We sketch the main sequence of events for this scenario in \Fig{fig:reheating}.

The CMB constraint on DM isocurvature matter perturbations can thus be expressed as an upper bound on the DM energy density sourced by the $s_0$ condensate as \cite{Kainulainen:2016vzv}
\begin{equation}
\label{isocurvature}
\frac{\rho_{\rm S}(T_{\rm CMB})}{\rho_{\rm A}(T_{\rm CMB})} \simeq \sqrt{\frac{\beta}{1-\beta}}\sqrt{\frac{\mathcal{P}_\zeta}{\mathcal{P}_{\delta_{s_0}}}} \,,
\end{equation}
where $\rho_{\rm S}$ and $\rho_{\rm A}$ are the isocurvature and adiabatic contributions to the DM density evaluated at last scattering of the CMB at $T_{\rm CMB} \simeq 0.3$ eV ---  that is, in our case, the $s$ particle DM sourced by the primordial $s_0$ condensate and Higgs decays, respectively. In this expression, $\mathcal{P}_\zeta \simeq 2.2\times 10^{-9}$ is the primordial curvature power spectrum \cite{Ade:2015xua}, $\mathcal{P}_{\delta_{s_0}}\equiv \langle (\delta\rho_{s_0}/\rho_{s_0})^2\rangle$ is the primordial power spectrum of $s_0$ fluctuations and $\beta\leq 0.05$ is the isocurvature parameter constrained by the \emph{Planck} data \cite{Ade:2015lrj}. We require that the Higgs decays into $s$ particles dominate over the DM yield from the primordial $s_0$ condensate. 

By assuming that the comoving number densities of the singlet scalars produced by the decay of the primordial $s_0$ condensate and Higgs decays are separately conserved, and together constitute all of the observed DM, $\rho_{\rm A,\mathrm{today}}/(3\Mp^2H^2_\mathrm{today})\simeq 0.12$, where $\Mp$ is the reduced Planck mass, one finds that \cite{Kainulainen:2016vzv}
\begin{equation}
\label{isocurvature_no_thermalisation}
\frac{\Omega_{\rm DM}^{(s_0)}h_{100}^2}{0.12} \simeq 
0.642\, \Omega_\gamma^{\frac34}h_{100}^\frac32 \lambda_s^{-\frac14} \frac{m_s}{\rm GeV}\left(\frac{s_*}{10^{11}{\rm GeV}}\right)^{\frac{3}{2}} \simeq \sqrt{\frac{\beta}{1-\beta}}\sqrt{\frac{\mathcal{P}_\zeta}{\mathcal{P}_{\delta_{s_0}}}}
 \,.
\end{equation}
In this expression, $h_{100} = H_{\rm today}/(100\,\mathrm{km}\, \mathrm{s}^{-1}\, \mathrm{Mpc}^{-1})$ parametrises the Hubble parameter today, $\Omega_\gamma$ is the dimensionless photon density parameter today and $s_*$ is the spectator field value during the last 60 \efolds of inflation (where it remains effectively constant).

By then using the typical value for $s_*$ given by \Eq{eq:stationary_s}, one obtains
\begin{equation}
\label{Pdeltas}
\mathcal{P}_{\delta_{s_0}} = \frac{9}{4}\frac{H_*^2}{(2\pi)^2s_*^2} \simeq  \left[ \frac{27\lambda_s}{128 \pi^2}\dfrac{\Gamma^2\left(\frac{1}{4}\right)}{\Gamma^2\left(\frac{3}{4}\right)} \right]^{\frac{1}{2}}  \,,
\end{equation}
and we can determine the Hubble scale to be
\begin{align}
\frac{H_*}{10^{11}{\rm GeV}} &\simeq 
4.89 \frac{\calP_\zeta^{\frac13}}{h_{100}\Omega_\gamma^\frac12} \left(\frac{\beta}{1-\beta}\right)^{\frac{1}{3}} \lambda_s^{\frac{1}{4}}\left(\frac{m_s}{{\rm GeV}}\right)^{-\frac{2}{3}} \nonumber \\
& \simeq 0.97 \left(\frac{\beta}{1-\beta}\right)^{\frac{1}{3}} \lambda_s^{\frac{1}{4}}\left(\frac{m_s}{{\rm GeV}}\right)^{-\frac{2}{3}} \label{H*_no_thermalisation}\,,
\end{align}
given our previously outlined assumptions, where for the second line we have used ${\cal P}_\zeta = 2.2 \times 10^{-9}$, $h_{100} =  0.673$ and $\Omega_\gamma = 9.3 \times 10^{-5}$~\cite{Ade:2015xua}. This result for $H_*$ then allows one to determine the energy scale of inflation independent of the inflationary tensor perturbations.

\begin{figure}[h!]
\begin{center}
\includegraphics[width=0.8\textwidth]{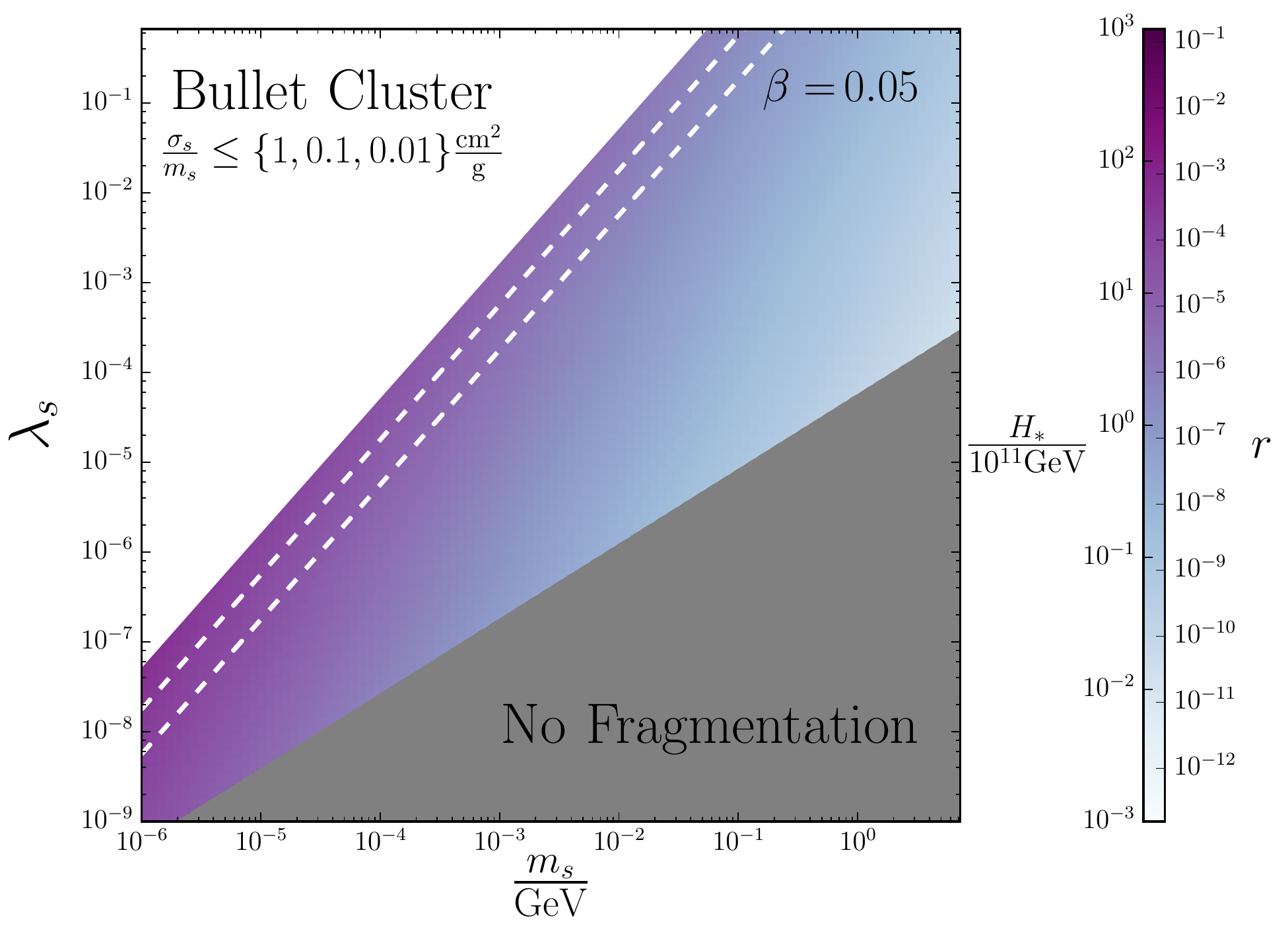}
\caption{\label{fig:H_*results} The value of the inflationary energy scale $H_*$ as a function of the singlet scalar mass $m_s$ and self-interaction strength $\lambda_s$, where values have been fixed using \Eq{H*_no_thermalisation} and setting the DM isocurvature relative amplitude to the \emph{Planck}~\cite{Ade:2015lrj} upper limit $\beta = 0.05$ for demonstration. The white region in the top left-hand corner represents the constraint on the self-interaction cross-section provided by the Bullet Cluster in \Eq{scrosssection}, with the dotted white lines indicating how this constraint strengthens with decreasing upper limits on $\sigma_s/m_s$. The grey region in the bottom right-hand corner is a consistency bound related to the requirement of fragmentation occurring in the quartic potential, \Eq{eq:consistency-frag-before-nrel}, which is computed in \App{app:general-scaling-energy-density}. In this figure, $\bar{w}_{\rm osc}=1/3$. 
}
\end{center}
\end{figure}

The value obtained for $H_*$, and the corresponding value for the tensor-to-scalar ratio $r$, are shown in \Fig{fig:H_*results}. The constraints on the DM self-interaction cross-section from observations of small-scale structure, namely the Bullet Cluster, have been superimposed. Indeed, in the limit where the singlet mass is much smaller than the Higgs mass, $m_s\ll m_h$, the singlet scalar self-interaction cross-section divided by its mass is given by~\cite{Kaplinghat:2015aga}
\begin{equation}
\label{scrosssection}
\frac{\sigma_s}{m_s} = \frac{9\lambda_s^2}{32 \pi m_s^3} \leq 1
 \frac{{\rm cm}^2}{{\rm g}}\,,
\end{equation}
where the upper bound applies when the $s$ particles constitute all DM. The exclusion zone that would be obtained from more stringent constraints on $\sigma_s/m_s$ is also displayed, in order to assess how parameter space could be even more reduced by improving the constraints on, or by measuring, the DM self-interaction cross-section. The result is not displayed in the grey region either, since it corresponds to values of the parameters for which fragmentation does not occur in the part of the potential dominated by the quartic term, and our calculation does not apply.

As shown in \App{app:general-scaling-energy-density} the result is valid for $\lambda_s\ll 1$, and for $m_s\gtrsim \mathcal{O}(1)$ keV because, otherwise, the $s$ particle DM is too hot and suppresses structure formation \cite{Murgia:2017lwo}. As discussed above, we also require $\lambda_{hs} < 10^{-7}$, as otherwise the singlet sector would thermalise with the SM sector and the primordial isocurvature perturbations would be washed away, $\beta=0$. Furthermore, despite the fact that the primordial singlet condensate yields only a subdominant contribution to the total DM abundance, the SM particle decays and annihilations can produce the rest of the DM abundance. This amounts to choosing a sufficiently large value for $\lambda_{hs}$, which for $m_s \in [10^{-6},1]$ GeV is roughly $\lambda_{hs} \in [10^{-12},10^{-9}]$ \cite{Yaguna:2011qn, Kang:2015aqa,Nurmi:2015ema,Bernal:2015xba,Kainulainen:2016vzv, Heikinheimo:2016yds,Heikinheimo:2017ofk}. The exact value, however, is not relevant for the minimal scenario (but will be in the extended one).

As discussed in \Refs{Nurmi:2015ema,Kainulainen:2016vzv,Heikinheimo:2016yds}, the above result for $H_*$ in \Eq{H*_no_thermalisation} is a generic consequence of a model where the additional scalar field is light and energetically subdominant during inflation and does not thermalise with the SM radiation after it. The result, however, is subject to a number of uncertainties related to dynamics in the inflaton sector, reheating history, and low-energy dynamics. We carefully consider these in the next section.

\section{Extended scenario}
\label{extended_scenario}

Due to uncertainties in the inflationary dynamics, reheating history, and low-energy dynamics, relaxing one or several assumptions we made above introduces modifications to our result \eqref{H*_no_thermalisation}. For example, even in slow-roll inflation, the Hubble rate may have a finite time dependence during inflation and the typical field displacement acquired by the spectator field at the end of inflation may change, or reheating might have taken a finite time, which introduces an arbitrary expansion history during which the primordial $s_0$ condensate grows its energy density with respect to the background, leading to different DM abundance today. 

These modifications can be effectively parameterised in \Eq{H*_no_thermalisation} as 
\begin{equation}
\label{H*_with_corrections}
\frac{H_*}{10^{11}{\rm GeV}} \simeq 0.97 \left(\frac{\beta}{1-\beta}\right)^{\frac{1}{3}} \lambda_s^{\frac{1}{4}}\left(\frac{m_s}{{\rm GeV}}\right)^{-\frac{2}{3}} \times \mu_{\rm inf}  \times \mu_{\rm reh} \times \mu_{\rm low} \,,
\end{equation}
where $\mu_{\rm inf}$, $\mu_{\rm reh}$, $\mu_{\rm low}$ are effective correction coefficients induced by inflationary dynamics, reheating history, and low-energy dynamics, respectively. Their detailed effect will be discussed one by one in the following subsections.

\subsection{Varying the inflationary dynamics} \label{sec:vary-inflation}

In this section we quantify the degree to which a finite time dependence of the Hubble rate during the early stages of inflation, \eg due to large-field corrections, $V\propto \phi^p$, to the inflaton potential, may affect our result. As was shown in \Ref{Hardwick:2017fjo}, the variance of $s$ at the end of inflation can be significantly larger than that given by \Eq{eq:stationary_s} depending on whether the distribution for $s$ has sufficient time to relax to the equilibrium distribution for a fixed value of $H$ --- the ``adiabatic'' regime in our terminology --- or whether, instead, the Hubble rate varies too fast (while still being in the slow-roll regime) for the system to relax to the equilibrium distribution. This can lead to a larger value for the variance than would be expected for a given, constant value of $H$. 

\begin{figure}[h!]
\begin{center}
\includegraphics[width=0.7\textwidth]{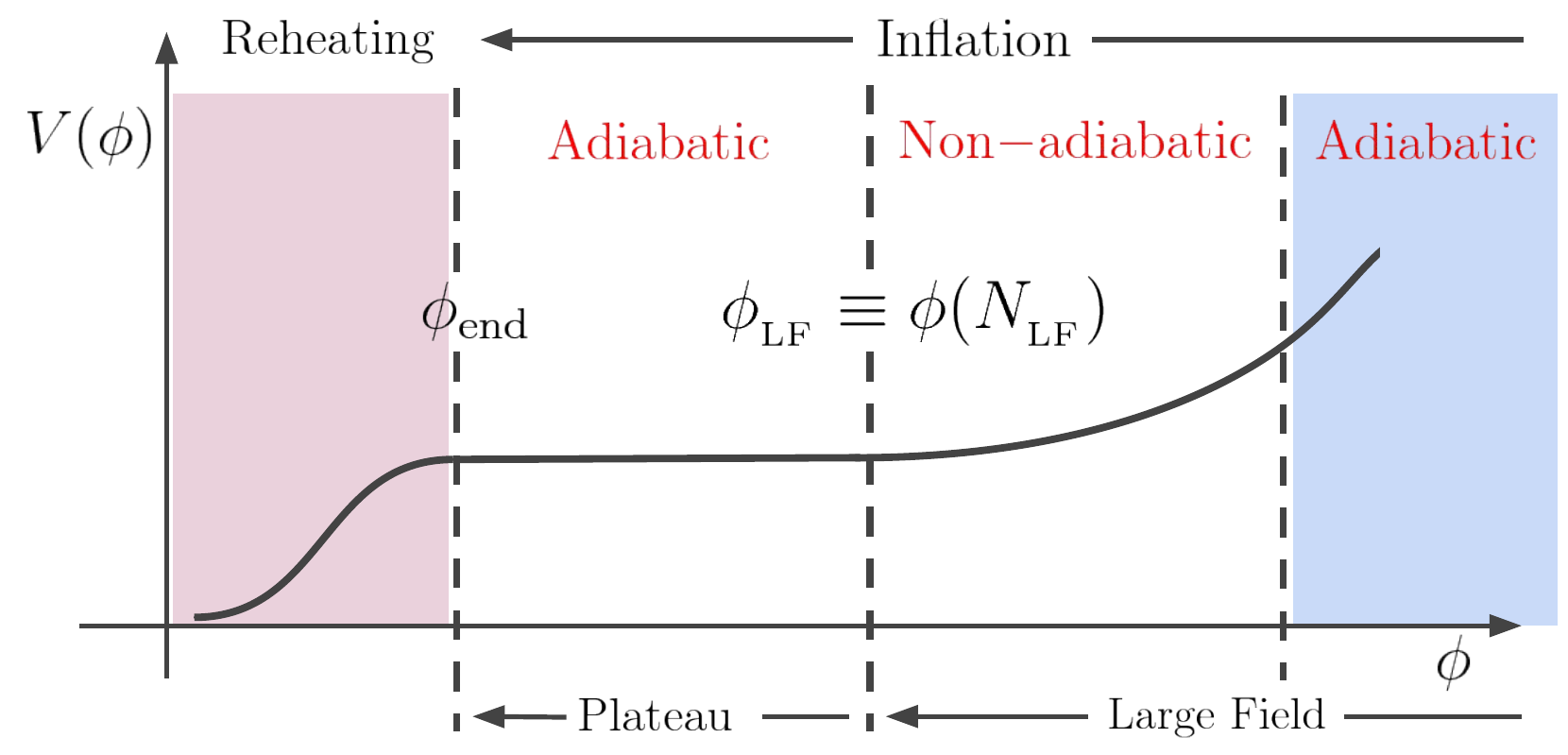}
\caption{\label{fig:potential} The inflationary plateau potential under large-field corrections studied in \Sec{sec:vary-inflation}. ``Adiabatic'' and ``non-adiabatic'' refer to different regimes of the stochastic dynamics of the $s$ field as described in the text.}
\end{center}
\end{figure}

In order to illustrate this effect, we shall consider a potential for the inflaton field $\phi$ which interpolates between a plateau potential (for explicitness here, we consider the Starobinsky model~\cite{Starobinsky:1980te}), consistent with \emph{Planck} constraints on the inflaton potential when observable scales leave the Hubble radius \cite{Martin:2013nzq, Ade:2015xua}, and a large-field model at early times when $\phi > \phi_{{}_{\rm LF}}$, such that
\begin{equation} \label{eq:potential}
V \left(\phi \right) = M^4 \left[ \left( 1- \ee^{-\sqrt{\frac{2}{3}} \frac{\phi}{\Mp}} \right)^2 + \left( \frac{\phi}{\phi_{{}_{\rm LF}}}\right)^p \right] \,.
\end{equation}
In this expression, $\phi_{{}_{\rm LF}} \gg  \Mp$ in order for the potential to be of the plateau type when observable scales leave the Hubble radius. In what follows, we assume $p>2$ since this gives the greatest correction to our result. The potential~(\ref{eq:potential}) is sketched in \Fig{fig:potential}. 

In both regimes we can identify the number of \efolds associated with two characteristic timescales: the relaxation timescale for the $s$ field to relax to the equilibrium distribution for a quartic potential, $N_{\rm relax} = 1/\sqrt{\lambda_s}$~\cite{Enqvist:2012xn, Hardwick:2017fjo}, and the timescale associated with a variation in the Hubble parameter, $N_H = H/(\dd H/\dd N) =  1/\epsilon $, where $\epsilon$ is the usual first slow-roll parameter. Using these timescales, the effect of the inflationary background evolution (i.e., the inflaton field rolling down in the potential~(\ref{eq:potential})) on the variance of $s$ can be divided into three phases: 
\begin{enumerate}
\item{At early times in the large-field regime, $\epsilon \sim (\Mp/\phi)^2$, and the $s$ field evolves adiabatically (hence the far-right label in \Fig{fig:potential}) because its relaxation timescale is shorter than the timescale associated with the variation of the Hubble parameter of the background, $N_{\rm relax}\ll N_H$.}
\item{Still within the large field regime, the value of $\epsilon$ gradually increases and $N_H$ decreases over time until $N_H < N_{\rm relax}$, at which point the evolution of $s$ ceases to be adiabatic and its variance effectively freezes in until the end of this phase~\cite{Hardwick:2017fjo} with a value we label $\langle s^2_{{}_{\rm LF}}\rangle$. }
\item{After the large-field regime ends, $N_H$ increases such that the condition $N_H> N_{\rm relax}$ is quickly fulfilled again. The $s$ field then begins to relax to its new equilibrium distribution on the plateau, but starting with an initial variance $\langle s^2_{{}_{\rm LF}}\rangle$ determined by the preceding large-field regime.}
\end{enumerate}

At the end of the large-field regime, the spectator field $s$ acquires a typical field displacement given by~\cite{Hardwick:2017fjo}
\begin{equation} \label{eq:eq-dist}
\langle s^2_{{}_{\rm LF}}\rangle = 12\left( 1-\frac{2}{p}\right)\frac{\Gamma^2 \left( \frac{3}{4} \right)}{\Gamma^2 \left( \frac{1}{4} \right)} \frac{H_\uend^2}{\lambda_s} \,,
\end{equation}
where $p$ has been defined in \Eq{eq:potential} and $H_\uend$ denotes the value of the Hubble parameter at the end of inflation (which is of the same order as the one along the plateau). The number of \efolds that must be realised to reach the stationary distribution~(\ref{eq:stationary_s}) is given by $N_{\rm relax}=1/\sqrt{\lambda_s}$. Therefore for the equilibrium distribution \Eq{eq:stationary_s} to be valid, we require a large number of \efolds on the plateau, $N_{\rm plateau} \gg N_{\rm relax} = 1/\sqrt{\lambda_s}$.

The variance of the spectator field at the end of the plateau phase, subject to the initial condition set by \Eq{eq:eq-dist}, can be written as
\begin{equation}
\left\langle s^2 \right\rangle  =
\dfrac{\Gamma\left(\frac{3}{4}\right)}{\Gamma\left(\frac{1}{4}\right)}\sqrt{\frac{3H_\uend^4}{2\pi^2\lambda_s}} \times {\mu_{\rm inf}^{-6} \left( N_{\rm plateau} \right) }\,,
\label{eq:corrected-variance}
\end{equation}
where $\mu_{\mathrm{inf}}$ defines the correction to \Eq{eq:stationary_s} and is given by~\cite{Hardwick:2017qcw}
\begin{equation}
\label{eq:muinf}
\mu_{\rm inf} \left( N_{\rm plateau} \right) = \left( \tanh\left\lbrace \sqrt{\dfrac{3\lambda_s}{8}}\dfrac{\Gamma\left(\frac{1}{4}\right)}{8\pi\Gamma\left(\frac{3}{4}\right)}N_{\rm plateau} +\mathrm{atanh} \left[  \frac{p}{3p-6} \frac{\Gamma\left(\frac{1}{4}\right)}{\Gamma\left(\frac{3}{4}\right)} \sqrt{\frac{3\lambda_s}{32\pi^2}} \right] \right\rbrace \right)^{\frac{1}{6}} \,.
\end{equation}
Note that since $s_*$ (hence $\mu_{\rm inf}$) appears in both sides of the last equality in \Eq{isocurvature_no_thermalisation}, through $s_{\rm *}$ directly and through $\mathcal{P}_{\delta_{s_0}}$ indirectly, see \Eq{Pdeltas}, the power of $\mu_{\rm inf}$ in \Eq{eq:corrected-variance} indeed yields a factor $\mu_{\rm inf}$ in \Eq{H*_with_corrections}.

\begin{figure}[h!]
\begin{center}
\includegraphics[width=0.7\textwidth]{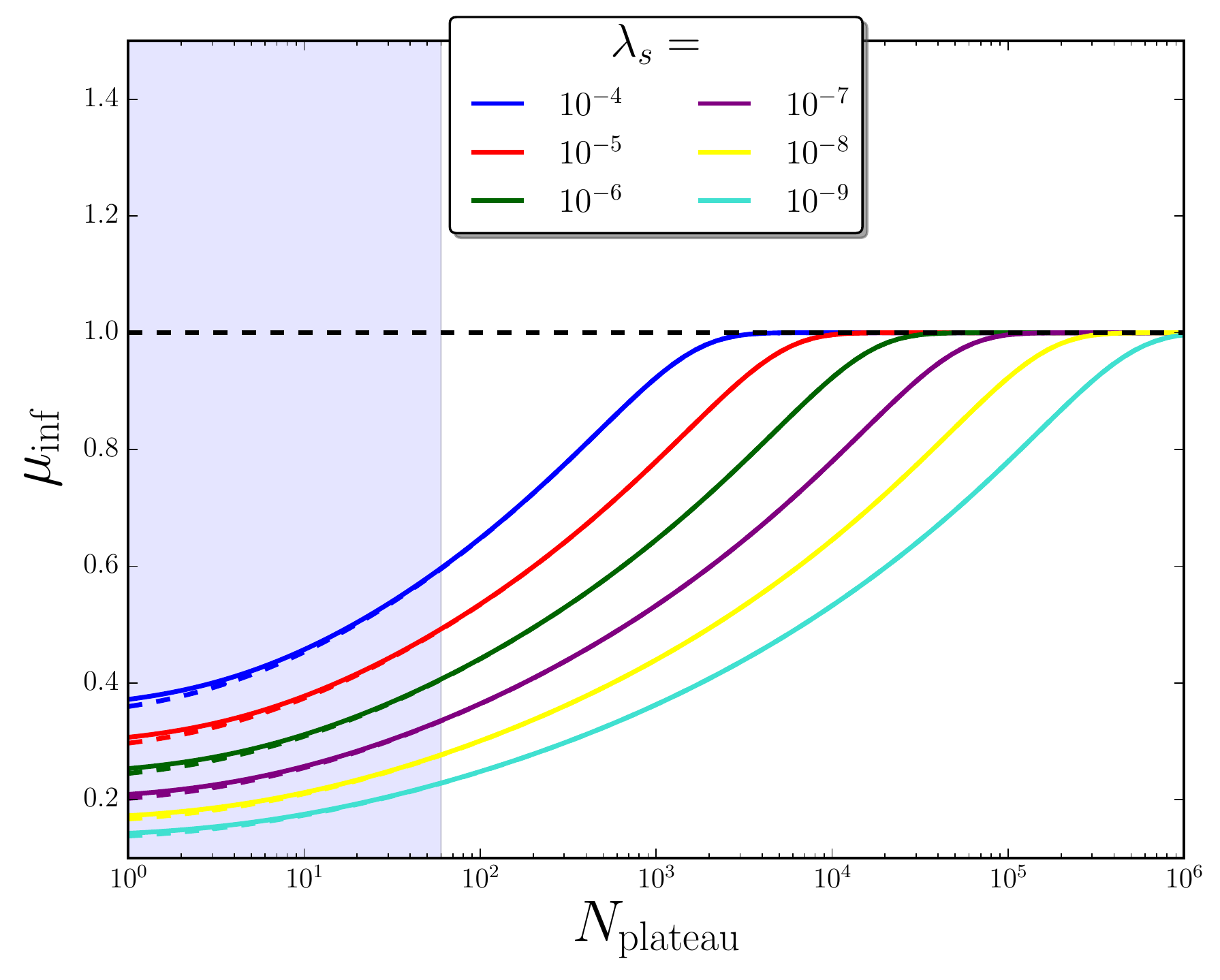}
\caption{\label{fig:Hend_over_Hadiab} Correction factor to the value of $H_*$ from the dynamics of the spectator field during the early stages of inflation, $\mu_{\mathrm{inf}}$, as a function of the number of \efolds $N_{\mathrm{plateau}}$ spent on the plateau in the potential depicted in \Fig{fig:potential}. The solid lines stand for $p=4$ in \Eq{eq:potential} and the dashed lines for $p=6$, which shows that the result is almost independent of $p$. The light blue shaded region corresponds to values of $N_{\rm plateau}$ that are too small to let the observable scales leave the Hubble radius in the plateau phase, as required from observations.
}
\end{center}
\end{figure}
\begin{figure}[h!]
\begin{center}
\includegraphics[width=0.8\textwidth]{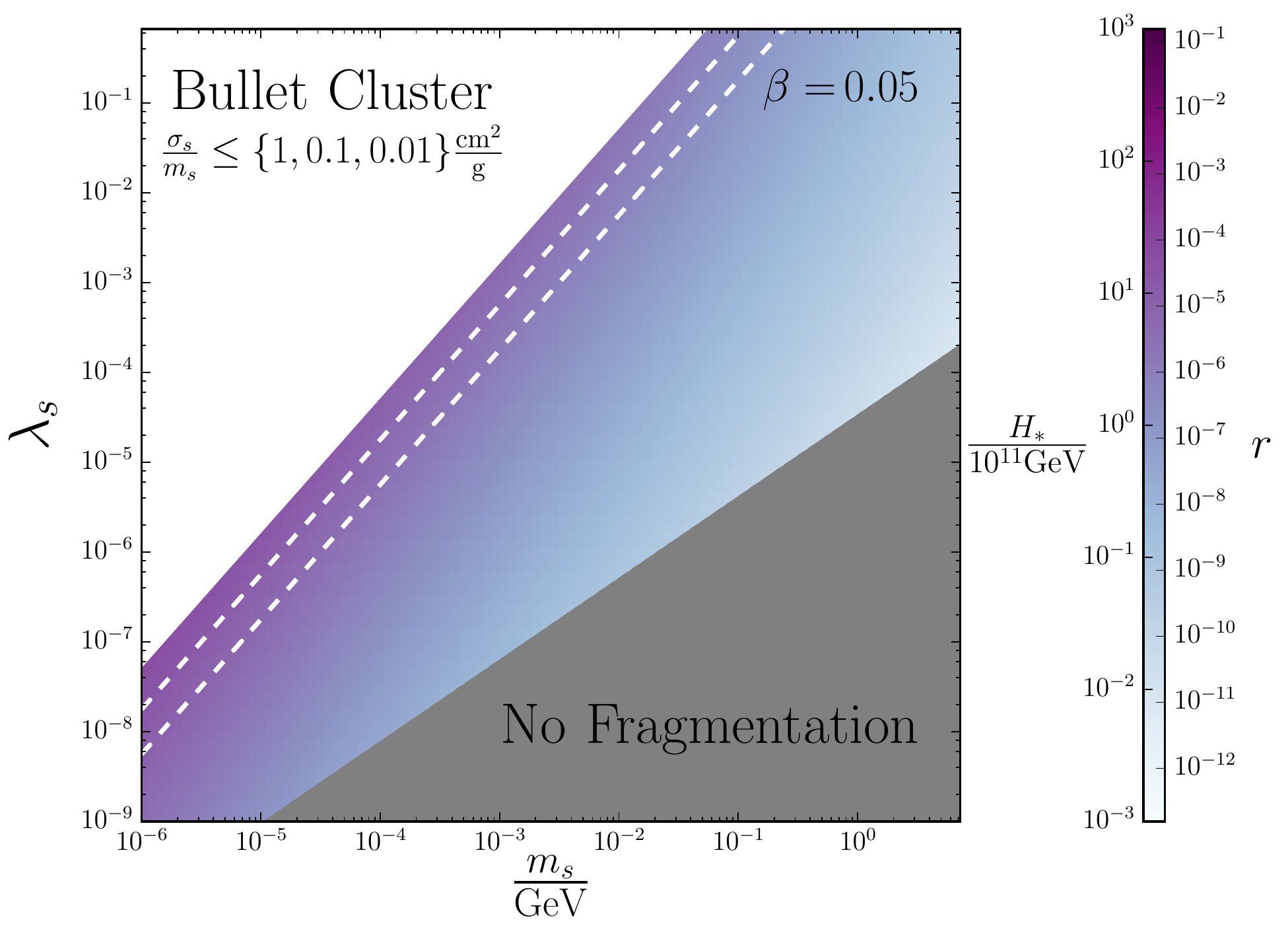}
\caption{\label{fig:H_*results_mu_inf} Same as \Fig{fig:H_*results}, but with the correction $\mu_{\rm inf}$ appearing in \Eq{H*_with_corrections} and defined in \Eq{eq:muinf} included, with $N_{\mathrm{plateau}}=100$ and $p=4$, yielding only small differences with \Fig{fig:H_*results}.
}
\end{center}
\end{figure}

The correction factor $\mu_{\rm inf}$ is displayed in \Fig{fig:Hend_over_Hadiab} as a function of the number of \efolds spent on the plateau, $N_{\mathrm{plateau}}$, for several values of $p$ and $\lambda_s$. One can check that when $N_{\mathrm{plateau}}$ is sufficiently large, $\mu_{\rm inf} \simeq 1$, and that the number of \efolds that need to be spent on the plateau in order to erase the imprint of the large-field early stage decreases with $\lambda_s$, in agreement with the formula $N_{\mathrm{relax}}=1/\sqrt{\lambda_s}$ given above. The result is almost independent of $p$. Since at least $\sim 60$ \efolds must be realised on the plateau, one can check that $\mu_{\rm inf}$ is always of order one, so that the value of $H_*$ computed from \Eq{H*_with_corrections} is impacted by the large-field corrections to the spectator field dynamics by at most an ${\cal O}(1)$ constant. This is also illustrated in \Fig{fig:H_*results_mu_inf}, where $H_*$ is displayed as a function of $m_s$ and $\lambda_s$ taking $N_{\mathrm{plateau}}=100$, and where the differences with \Fig{fig:H_*results} are very mild.

Finally, let us check that, as assumed in the above calculation, during inflation the quartic term in the scalar potential dominates over the quadratic one, $\lambda_s \langle s^2\rangle \gg 2m_s^2+\lambda_{hs}\langle h^2\rangle$. Using \Eq{eq:corrected-variance} to estimate $\langle s^2 \rangle$ in the first condition $\lambda_s \langle s^2\rangle \gg 2m_s^2$, one obtains $H_* \gg m_s \lambda_s^{-1/4} \mu_{\mathrm{inf}}^3$. In all following figures, we make sure that this condition is always satisfied. Using a relation similar to \Eq{eq:corrected-variance} to estimate $\langle h^2 \rangle$ in the second condition $\lambda_s \langle s^2\rangle \gg \lambda_{hs}\langle h^2\rangle$, one obtains  $\sqrt{\lambda_h \lambda_s} \gg \lambda_{hs}$, where $\lambda_h$ is the self-interaction strength of the Higgs. As noted above, to prevent the singlet $s$ from thermalising with the SM in the early Universe,  one must have $\lambda_{hs}<10^{-7}$, and since we assume $\lambda_h \gtrsim 10^{-5}$~\cite{Figueroa:2015rqa}, the lower bound on $\lambda_s$ used in all figures is such that this condition is always satisfied too.

\subsection{Varying the reheating history} \label{sec:vary-reheating}

We now turn our attention to the second possible modification in \Eq{H*_with_corrections}, namely the reheating expansion history. So far we have assumed that after inflation, the energy density of the background decays as radiation. If this is not the case, the abundance of DM obtained from the particles into which the condensate fragments is different, hence the inferred value of $H_*$ changes. 

\begin{figure}[h!]
\begin{center}
\includegraphics[width=0.8\textwidth]{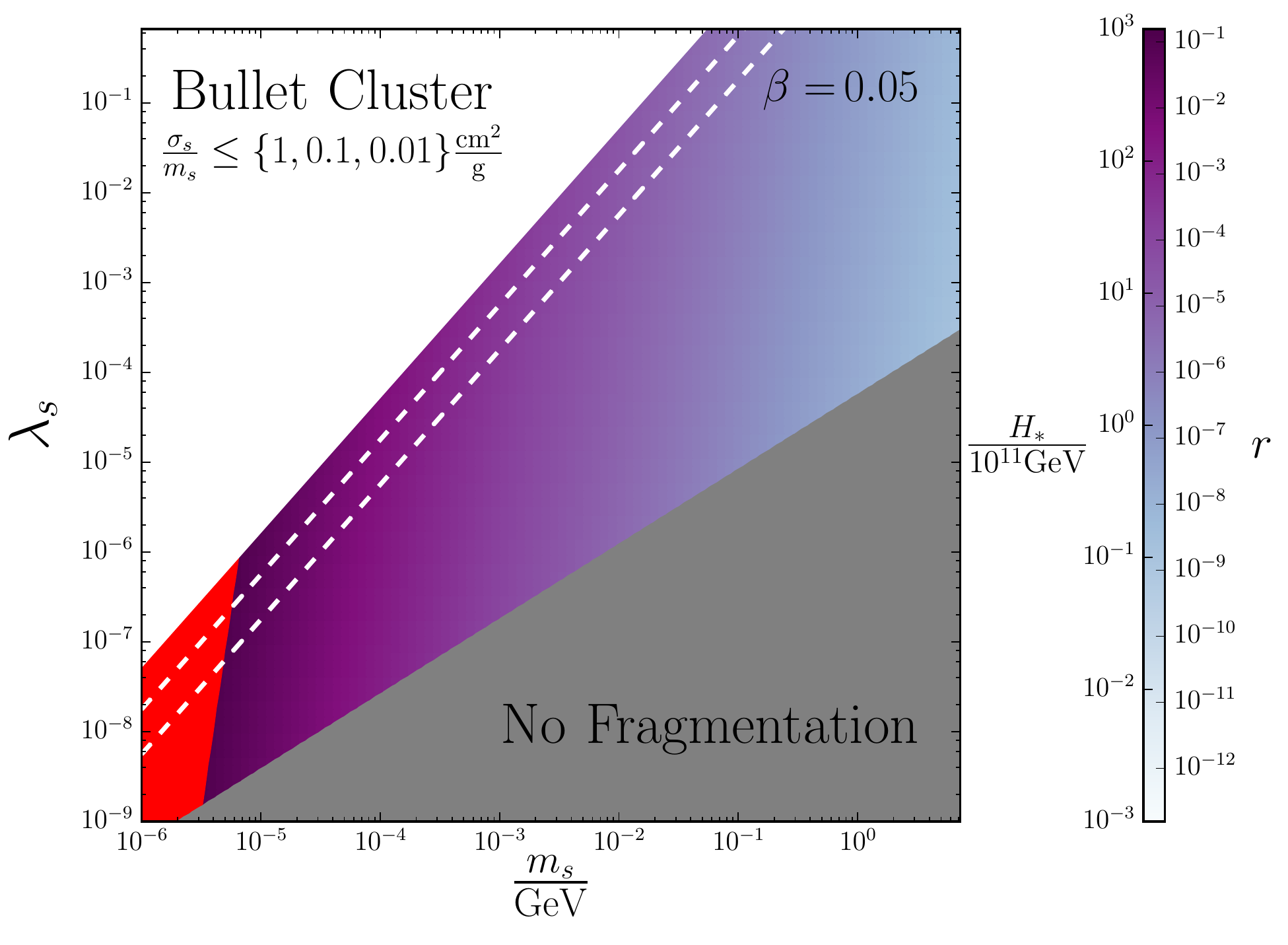}
\caption{\label{fig:H_*results_mu_reh} 
Same as \Fig{fig:H_*results}, but with the correction $\mu_{\rm reh}$ appearing in \Eq{H*_with_corrections} and defined in \Eq{eq:mu_reh}, included, with $\bar{w}_{\rm osc}=0.23$. The region shaded in red is ruled out since it yields values for the tensor-to-scalar ratio that are larger than the observational upper bound $r< 0.12$~\cite{Ade:2015tva}.}
\end{center}
\end{figure}

In \App{app:general-scaling-energy-density} we provide a detailed calculation of the energy density contained in the singlet particles at the end of the multi-stage process depicted in \Fig{fig:reheating}, for an arbitrary background expansion history between the end of inflation and the fragmentation time (for the result we use about fragmentation rate to apply, the Universe needs to be in a radiation era at the fragmentation time). We find that the result only depends on the average equation-of-state parameter during the oscillation phase of the condensate, $\bar{w}_{\mathrm{osc}}$, and on the quartic coupling constant $\lambda_s$. More precisely, an analogous expression to \Eq{isocurvature_no_thermalisation} is obtained,
\begin{align} 
\frac{\Omega^{(s_0)}_{\rm DM}h^2_{100}}{0.12} &=  0.642\, \Omega_\gamma^{\frac34}h_{100}^\frac32 \lambda_s^{-\frac14} \frac{m_s}{\rm GeV}\left(\frac{s_*}{10^{11}{\rm GeV}}\right)^{\frac{3}{2}} \times \mu_{\rm reh}^{-\frac{3}{2}}(\lambda_s,\bar{w}_{\rm osc})  \label{eq:mu_reh_obtain_1}\,,
\end{align}
where we have defined
\begin{equation} \label{eq:mu_reh}
\mu_{\rm reh}(\lambda_s,\bar{w}_{\rm osc}) \equiv \left( \frac{\alpha \lambda_s}{\sqrt{3}}\right)^{\frac{3\bar{w}_{\rm osc}-1}{3\bar{w}_{\rm osc}+1}} \,.
\end{equation}
One can check that the power to which $\mu_{\mathrm{reh}}$ appears in \Eq{eq:mu_reh_obtain_1} is such that it appears with power one in \Eq{H*_with_corrections}. In this expression, $\alpha \simeq 0.023$ is a numerical constant that comes from the calculation of the fragmentation rate. When $\bar{w}_{\rm osc}=1/3$, $\mu_{\mathrm{reh}}=1$ and \Eq{isocurvature_no_thermalisation} is recovered. In \App{app:general-scaling-energy-density}, we also derive and carefully study the conditions under which the assumptions made in the timeline of \Fig{fig:reheating} are satisfied. In particular, this results in the ``no fragmentation'' exclusion zone in \Figs{fig:H_*results}, \ref{fig:H_*results_mu_inf}, \ref{fig:H_*results_mu_reh} and \ref{fig:H_*results_mu_low}.

\begin{figure}[h!]
\begin{center}
\includegraphics[width=0.7\textwidth]{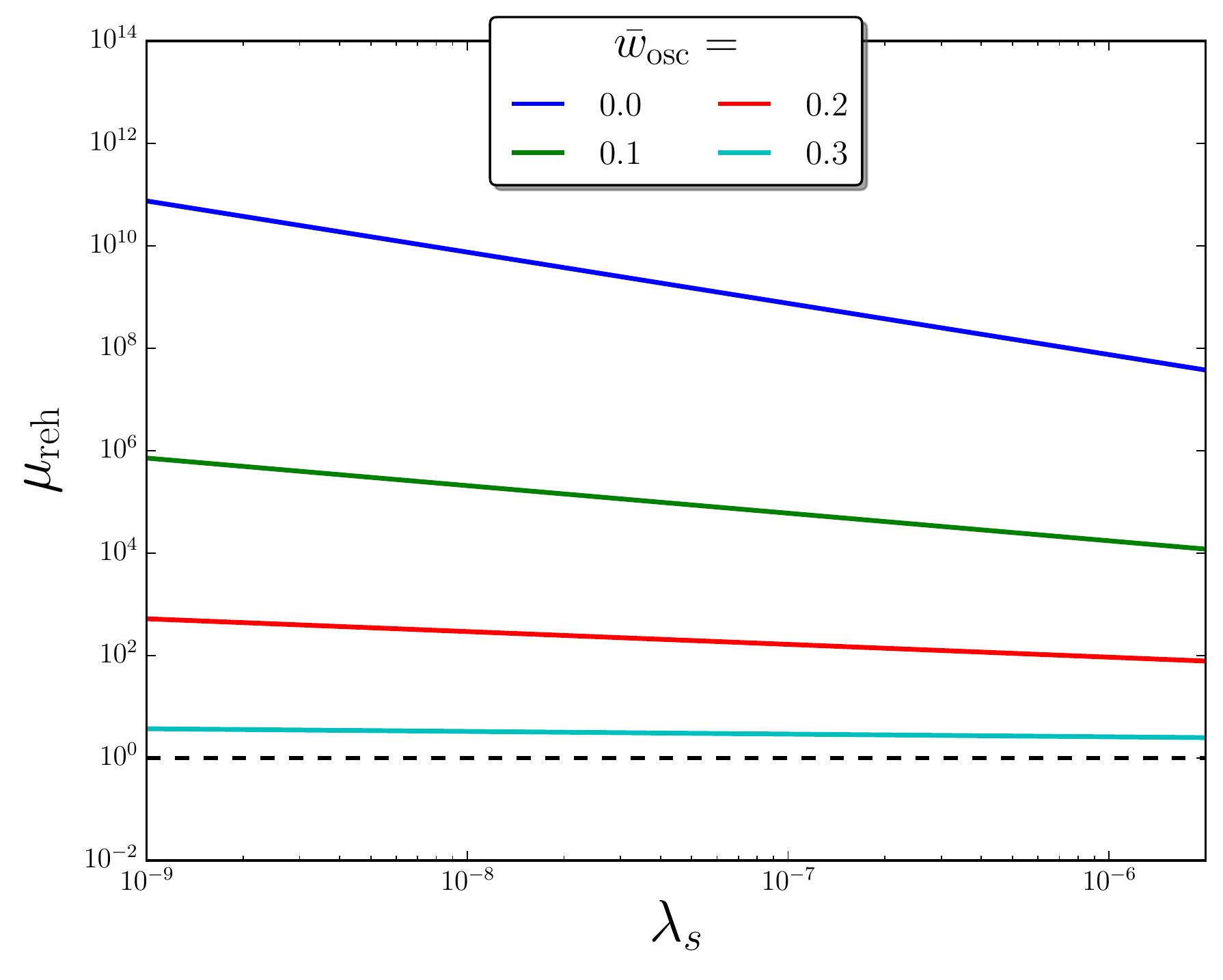}
\caption{\label{fig:mu_reh} Correction factor $\mu_{\mathrm{reh}}$ appearing in \Eq{H*_with_corrections} and accounting for an arbitrary expansion history during reheating, plotted as a function of the self-interaction strength of the singlet scalar $\lambda_s$ for a few values of the background average equation-of-state parameter $\bar{w}_{\mathrm{osc}}$ during the oscillation phase of the condensate after inflation.}
\end{center}
\end{figure}

The correction factor $\mu_{\mathrm{reh}}$ is plotted as a function of $\lambda_s$ for a few values of $\bar{w}_{\rm osc}$ in \Fig{fig:mu_reh}. Unlike the correction factor $\mu_{\rm inf}$ in the preceding subsection, we see that $\mu_{\rm reh}$ can vary by many orders of magnitude when $\bar{w}_{\rm osc}$ departs from $1/3$. This is also illustrated in \Fig{fig:H_*results_mu_reh}, where $H_*$ is displayed as a function of $m_s$ and $\lambda_s$ taking $\bar{w}_{\mathrm{osc}}=0.23$, and where the difference with \Fig{fig:H_*results} is quite large. There even are regions (in red) for which the predicted value of the tensor-to-scalar ratio is too large to satisfy observational bounds~\cite{Ade:2015tva}.

One notices that if $\bar{w}_{\mathrm{osc}}<1/3$, $\mu_{\mathrm{reh}}>1$ and the inferred value of $H_*$ in the minimal setup is smaller than the actual one, while if $\bar{w}_{\mathrm{osc}}>1/3$, $\mu_{\mathrm{reh}}<1$ and the inferred value of $H_*$ is larger than the actual one. The large effect from the reheating expansion history on our estimate of $H_*$ should be taken with a grain of salt since in practice, $\bar{w}_{\mathrm{osc}}$ may not depart too much from $1/3$. At the end of the oscillating phase indeed, one must have a background equation of state $w=1/3$ (for our expression for the fragmentation rate in \Eq{gamma} to apply), so $\bar{w}_{\mathrm{osc}}$ receives a contribution from values close to $1/3$. Let us also note that linear instabilities on small scales have been shown to yield $w=1/3$ very quickly after the end of inflation, in fact well before the inflaton field has effectively decayed~\cite{Lozanov:2016hid}. Such a mechanism would yield $\bar{w}_{\mathrm{osc}}=1/3$, leaving no imprint from the reheating expansion history on our result. 

\subsection{Low-energy dynamics} \label{sec:low-energy-dynamics}

In addition to corrections arising from dynamics during and immediately after inflation, there are corrections arising from particle dynamics at low energies, namely below the electroweak scale after the SM particle decays and annihilations have yielded the initial $s$ particle abundance. Contrary to the variations in the inflationary and the reheating dynamics studied in \Secs{sec:vary-inflation} and~\ref{sec:vary-reheating}, respectively, the low-energy dynamics effect is not a variation to the minimal setup but rather an inevitable correction that is inherent to it. It should therefore be understood as part of the minimal scenario.

\begin{figure}
\begin{center}
\includegraphics[width=.54\textwidth]{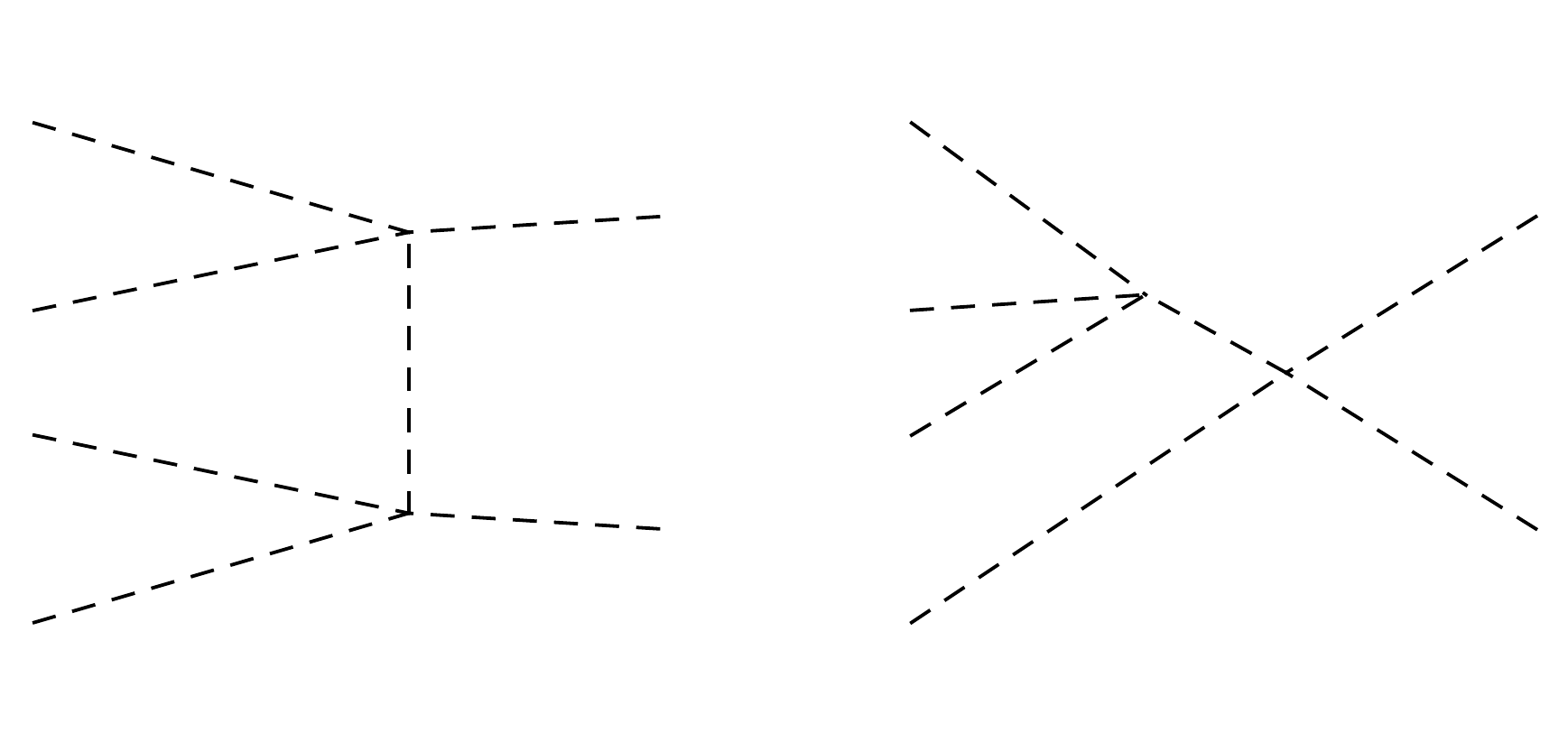}
\caption{Examples of relevant Feynman diagrams for the $4\to 2$ scalar self-annihilation process at the limit $m_s \ll m_h$.}
\label{diagram}
\end{center}
\end{figure}

Even though the portal coupling between DM and the SM sector is assumed to be so small that the $s$ particles never enter thermal equilibrium with the SM particles, it may happen that the $s$ particles reach chemical equilibrium within the singlet sector if the singlet sector has sufficient self-interactions. This leads to a characteristic hidden sector temperature $T_s$ different from the SM photon temperature $T$. If the singlet self-interactions are sufficiently strong, they can maintain the equilibrium for some time also after the singlet particles have become non-relativistic, leading to so-called \textit{DM cannibalism} \cite{Carlson:1992fn}, where number-changing interactions, such as $4\rightarrow 2$ annihilations\footnote{The $2\rightarrow 3$ annihilations are in our case forbidden due to the assumed $\mathbb{Z}_2$ symmetry of the scalar field.} (see \Fig{diagram}), reduce the singlet particle number density and heat the singlet sector with respect to the SM sector. Depending on the strength of singlet self-interactions, the $s$ number density can be significantly depleted before its final freeze-out from the equilibrium in the singlet sector. Thus the final DM abundance depends not only on the portal coupling $\lambda_{hs}$ and the mass $m_s$, but on a combination of the parameters $\lambda_{hs}$, $\lambda_s$ and $m_s$. This production mechanism is called \textit{dark} or \textit{hidden freeze-out}~\cite{Carlson:1992fn, Bernal:2015ova, Bernal:2015xba, Heikinheimo:2016yds, Bernal:2017mqb, Heikinheimo:2017ofk, Bernal:2017kxu}.

The main result \eqref{H*_no_thermalisation} applies only if there are no number-changing interactions in the singlet sector, \ie if the quartic scalar self-interaction strength $\lambda_s$ is small enough. The critical value above which the number-changing interactions play a significant role in determining the final DM abundance is \cite{Heikinheimo:2016yds}
\begin{equation}
\label{lambdas_fi}
\lambda_s^{(\rm fi)} \simeq \sqrt{\frac{19.4\left[g_*\left(m_h\right)g_*\left(m_s\right)\right]^\frac14 \sqrt{m_h m_s}}{\lambda_{hs}^{(\rm fi)} (m_s) \Mp }} 
\simeq 2.3 \times 10^{-8}\left[ \lambda_{hs}^{(\rm fi)}(m_s)\right]^{-\frac{1}{2}}\left( \frac{m_s}{{\rm GeV}}\right)^{\frac{1}{4}}\,,
\end{equation}
where $g_*(T)$ is the effective number of relativistic degrees of freedom in the SM plasma at temperature $T$ and $\lambda_{hs}^{(\rm fi)}(m_s)$ is the value of the portal coupling that yields the observed DM abundance for a given mass $m_s$ in the usual freeze-in case. For $\lambda_s<\lambda_s^{(\rm fi)}$ the usual freeze-in picture and the result \eqref{H*_no_thermalisation} are sufficient. Recalling that we assume $m_s \ll m_h$ in order for \Eq{scrosssection} to hold, the value of $\lambda_{hs}^{(\rm fi)}$ is determined by the usual freeze-in relation \cite{McDonald:2001vt,Hall:2009bx,Bernal:2017kxu}
\begin{equation}
\lambda_{hs}^{(\rm fi)} \simeq 10^{-11}\left(\frac{m_s}{\rm GeV}\right)^{-1/2}\,.
\end{equation}

If the number-changing interactions in the singlet sector become active, the $s$ particles equilibrate among themselves before the formation of the CMB. After the equilibration, the singlet scalar particles from both origins --- Higgs decays and primordial $s_0$ condensate fragmentation --- contribute to the thermal bath of DM, so that the relative abundance of the isocurvature component with respect to the total DM abundance remains constant from there on, as discussed in \Ref{Heikinheimo:2016yds}. %We assume that the thermalisation of the $s$ particles takes place at $T_{\rm therm}\simeq m_s$, which is the latest moment when the $s$ particles can reach chemical equilibrium with themselves.

The abundances from the primordial isocurvature condensate source $\rho^{(s_0)}_{\rm DM}(T_{\rm therm})$ and the adiabatic Higgs freeze-in source $\rho^{({\rm fi})}_{\rm DM}(T_{\rm therm})$ at the time of the thermalisation can be found by scaling the result in \Eq{isocurvature_no_thermalisation} by $a^3$ from the CMB temperature today, $T_{\rm today}\simeq 2.725$ K, up to $T=m_{ s}$, and scaling the usual freeze-in abundance of scalars by $a^{-4}$ down to the same temperature.\footnote{This is due to the fact that above $T=m_s$ both $\rho_{\rm DM}^{(s_0)}$ and $\rho_{\rm DM}^{({\rm fi})}$ scale as $a^{-4}$ \cite{Nurmi:2015ema,Heikinheimo:2016yds}, and thus their ratio at a thermalisation temperature $T_{\rm therm} > m_s$ would be the same as it is at $T_{\rm therm} = m_s$. Thus, it suffices to evaluate the ratio at $T_{\rm therm} = m_s$, which is the latest moment when the $s$ particles can reach chemical equilibrium with themselves. Moreover, reminiscent to the standard WIMP case, the final DM abundance only depends on the time of the freeze-out. Therefore the scenario is not sensitive to the exact temperature at which the hidden sector thermalisation occurs.} The isocurvature abundance is \cite{Kainulainen:2016vzv}
\begin{equation}
\label{rhoS}
\frac{\rho^{(s_0)}_{\rm DM}(T_{\rm therm})}{\GeV^4} \simeq 2.97 \times 10^{-52}\lambda_{s}^{-\frac{1}{4}}
\frac{m_{ s}}{{\rm GeV}}
\left(\frac{s_*}{10^{11}{\rm GeV}} \right)^{\frac{3}{2}}
\frac{g_{*{\cal S}}(m_{ s})}{g_{*{\cal S}}(T_{\rm today})} 
\left(\frac{m_{ s}}{T_{\rm today}}\right)^3 \,,
\end{equation}
where $g_{*{\cal S}}=g_{*{\cal S}}(T)$ is the effective number of entropy degrees of freedom in the radiation heat bath, and the usual freeze-in abundance of scalars is~\cite{Heikinheimo:2016yds}
\begin{eqnarray}
\label{rhoA}
\rho^{(\rm fi)}_{\rm DM}(T_{\rm therm})  &\simeq& m_{ s}n_{ s}(m_{ h})
\left[\frac{a(m_{h})}{a(m_{ s})} \right]^3 \\ \nonumber
&\simeq& \frac{3m_{ s}n_{h}^{\rm eq}(m_h)\Gamma_{h\to ss}}{H(m_{ h})}\frac{g_{*{\cal S}}(m_{ s})}{g_{*{\cal S}}(m_{ h})}\left(\frac{m_{ s}}{m_{ h}}\right)^3 \\ \nonumber
&=& \frac{3e^{-1}}{(2\pi)^{\frac{3}{2}}}\sqrt{\frac{90}{\pi^2}}\frac{g_{*{\cal S}}(m_{ s})}{g_{*{\cal S}}(m_{ h})\sqrt{g_{*}(m_{ h})}}\frac{\Gamma_{h \to ss}\Mp m_{ s}^4}{m_{ h}^2}\,,
\end{eqnarray}
where $n_s$ and $n_h^{\rm eq}$ are the singlet scalar and Higgs number densities, respectively, $a(m_i)$ is the scale factor at the time the photon temperature is $T=m_i, i=h,s$, and where in the limit $m_{ s}\ll m_{ h}$,
\begin{equation}
\label{gamma}
\Gamma_{h\rightarrow ss} = \frac{\lambda_{ hs}^2 v^2}{32\pi m_{ h}} \, .
\end{equation}
To derive this expression, we have assumed that the $h$ particles obey Maxwell-Boltzmann statistics after the electroweak symmetry breaking and that the singlet scalars are produced by $h\to ss$ at $T=m_{h}$.%, and that the thermalisation of scalars takes place no earlier than $T=m_{ s}$.

\begin{figure}[h!]
\begin{center}
\includegraphics[width=0.7\textwidth]{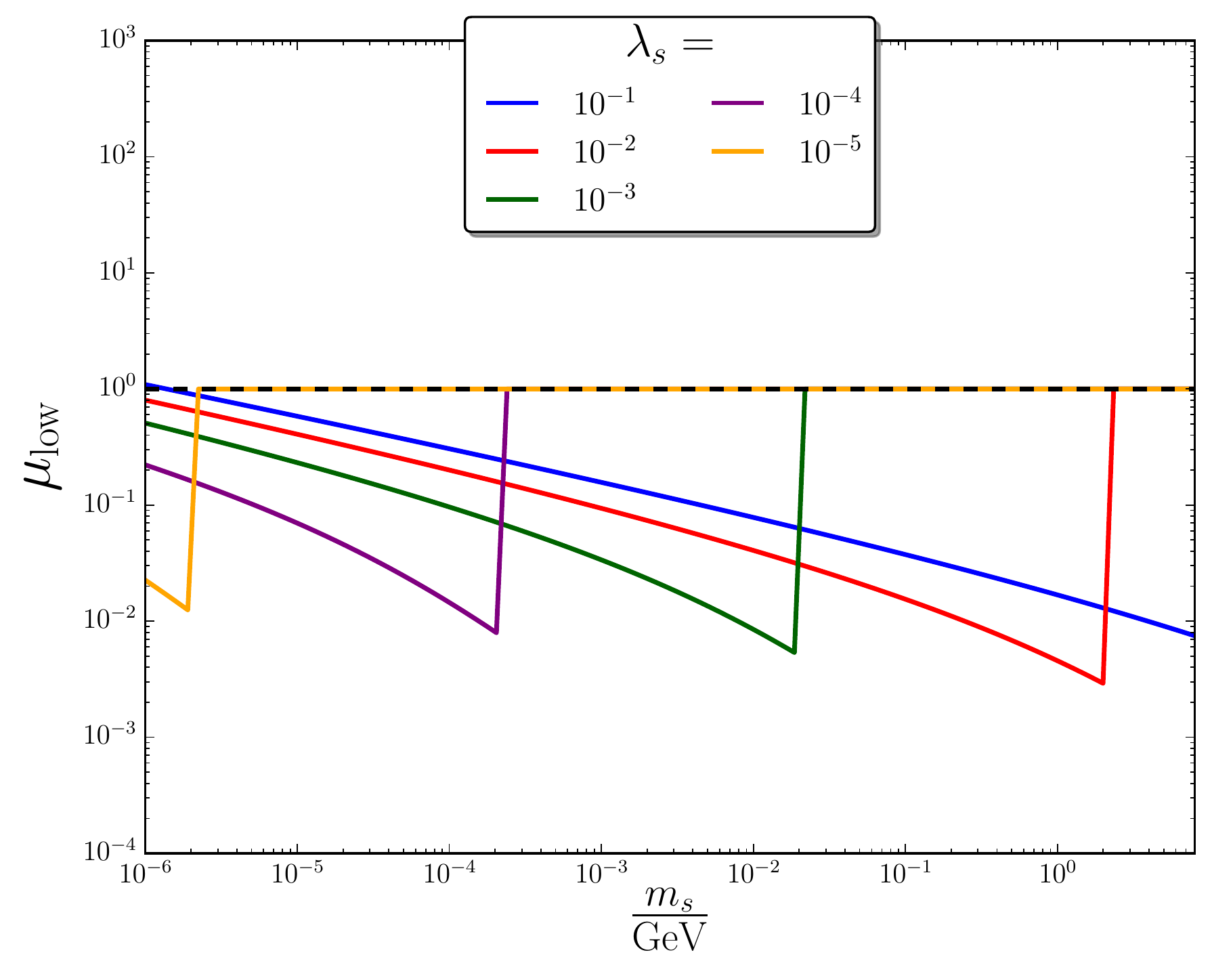}
\caption{\label{fig:mu_low} Correction factor $\mu_{\mathrm{low}}$ appearing in \Eq{H*_with_corrections} and defined in \Eq{mulow:2}, accounting for variations in the low-energy dynamics, plotted as a function of the singlet mass $m_s$ for a few values of the self-interaction strength $\lambda_s$. The vertical jumps correspond to critical values of $m_s$ above which $\lambda_s^{(\mathrm{fi})}>\lambda_s$, where $\lambda_s^{(\mathrm{fi})}$ is given in \Eq{lambdas_fi}. In such a case the usual freeze-in picture applies and  $\mu_{\rm low}= 1$.}
\end{center}
\end{figure}

Plugging \Eqs{rhoS} and~(\ref{rhoA}) into \Eq{isocurvature}, and using \Eq{Pdeltas} as before, we then obtain
\begin{equation}
\label{H*_thermalisation}
\frac{H_*}{10^{11}{\rm GeV}} \simeq 0.97\left(\frac{\beta}{1-\beta}\right)^{\frac13}\lambda_s^{\frac14}\left(\frac{m_s}{{\rm GeV}}\right)^{-\frac23}\times \mu_{\rm low}(\lambda_{hs},m_s) ,
\end{equation}
where we have defined
\begin{equation}
\label{mulow:1}
\mu_{\rm low}(\lambda_{hs},m_s) \equiv 8\times 10^{13}\lambda_{hs}^{\frac43}\left(\frac{m_s}{{\rm GeV}}\right)^{\frac23} .
\end{equation}
Because the result now depends explicitly on $\lambda_{hs}$, its exact value becomes important. As discussed above, we require that the singlet particles constitute all DM, which allows us to fix $\lambda_{hs}$ in terms of $\lambda_{s}$ and $m_{s}$, as shown in \App{relating_lambdahs_to_lm}, see \Eq{eq:lambdahs:ms:lambdas}. This gives rise to
\begin{equation}
\label{mulow:2}
\mu_{\rm low}(\lambda_{s},m_s) = 4\times 10^{-3}
W_0^{\frac{8}{9}} \left[7.1\times 10^{4} \lambda_s^{\frac{24}{11}}\left( \frac{m_s}{{\rm GeV}}\right)^{-\frac{10}{11}} \right]
\left(\frac{m_s}{{\rm GeV}}\right)^{-\frac29} .
\end{equation}
The correction factor $\mu_{\rm low}$ is plotted in \Fig{fig:mu_low}, where one can see that the reduction in $H_*$ caused by variations in the low-energy dynamics is at most of order $\order{10^{-3}}$.

\begin{figure}[h!]
\begin{center}
\includegraphics[width=0.8\textwidth]{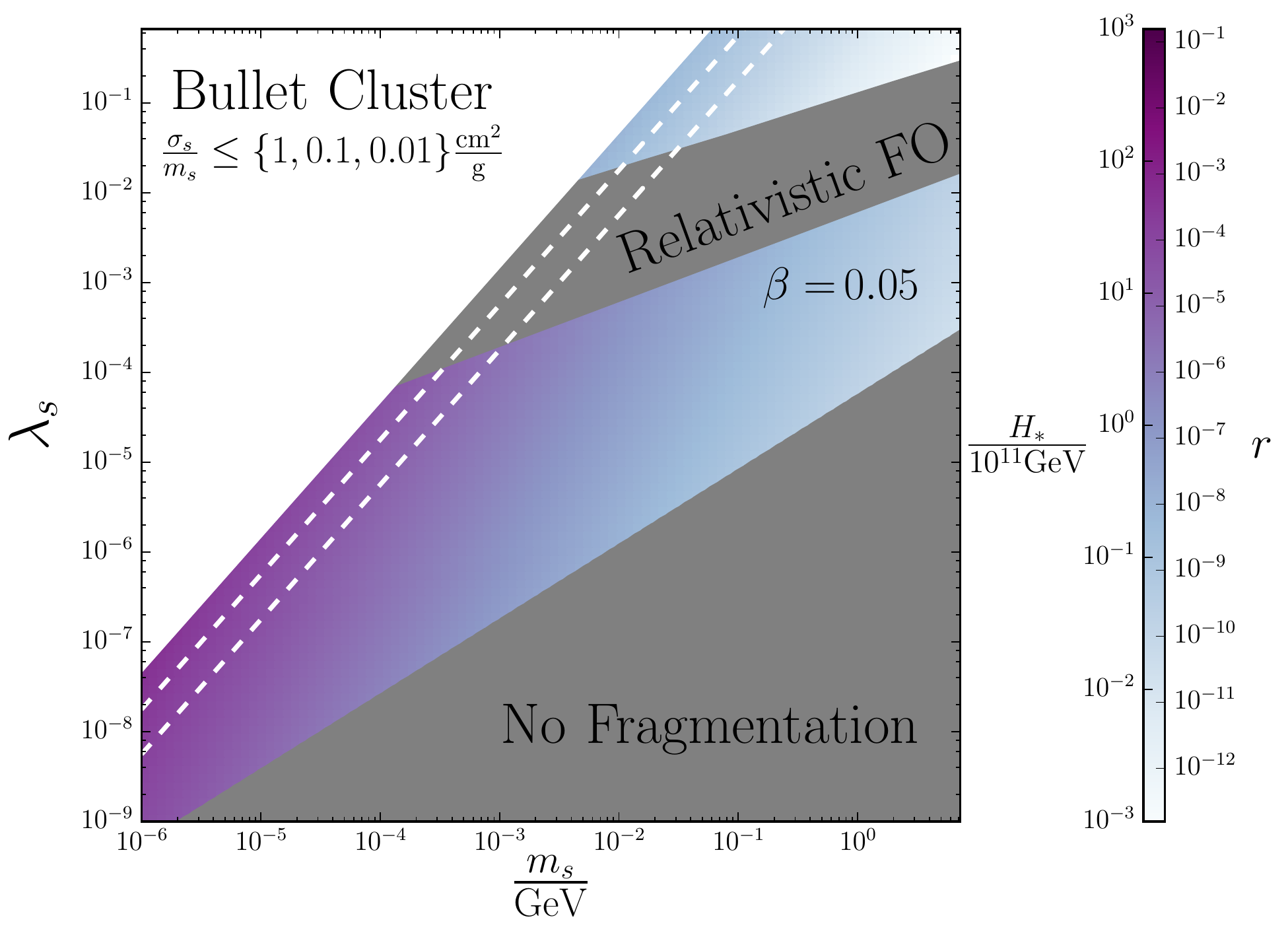}
\caption{\label{fig:H_*results_mu_low} Same as \Fig{fig:H_*results}, but with the correction $\mu_{\rm low}$ appearing in \Eq{H*_with_corrections} and defined in \Eq{mulow:2} included. In the grey region labeled ``relativistic FO'' for ``relativistic freeze-out'' the dark freeze-out occurs while the singlet particles are still relativistic and the present calculation does not apply. Above this grey region, the singlet particles thermalise within themselves and the DM abundance is determined by dark freeze-out at $T\lesssim m_s/3$ instead of the usual freeze-in mechanism at $T\simeq m_h$. Below this grey region, the DM abundance is determined by the usual freeze-in mechanism, $\mu_{\rm low} = 1$, and one retains the results of \Fig{fig:H_*results}.}
\end{center}
\end{figure}

The corresponding effect on $H_*$ is shown in \Fig{fig:H_*results_mu_low}. In the upper grey region shown in \Fig{fig:H_*results_mu_low} the dark freeze-out occurs while the singlet particles are still (semi-)relativistic, $m_s/T_s^{({\rm fo})}\leq 3$, and finding a solution that yields the correct DM abundance in that region requires a detailed numerical analysis, as discussed in \Ref{Heikinheimo:2016yds}. We will postpone that for future work. Above the upper grey region the freeze-out occurs at temperatures where the DM is non-relativistic, $m_s/T_s^{({\rm fo})}> 3$, and below this region the singlet particles do not thermalise within the singlet sector and the usual freeze-in picture is sufficient. In general, thermalisation of the singlet sector {\it increases} the number density of the $s$ particles, resulting in a larger final DM abundance than in the standard freeze-in scenario \cite{Heikinheimo:2016yds,Bernal:2015xba,Bernal:2015ova}, and in order to produce the observed DM abundance, a smaller initial abundance sourced by the SM particles is needed. Thus, an initial population of scalars produced from the decay of the primordial $s_0$ condensate contributes a larger fraction of the total DM energy density than it would in the standard freeze-in scenario, and hence the isocurvature contribution is larger. Thus, to keep the ratio \eqref{isocurvature} constant for fixed $\beta$, a smaller value for $\rho_{\rm S}$, \ie a smaller value for $H_*$, is needed. This explains why the correction factor $\mu_{\rm low}$ is always less than $1$.
\section{Conclusions} \label{sec:conclusions}
\begin{figure}[h!]
\begin{center}
\includegraphics[width=0.74\textwidth]{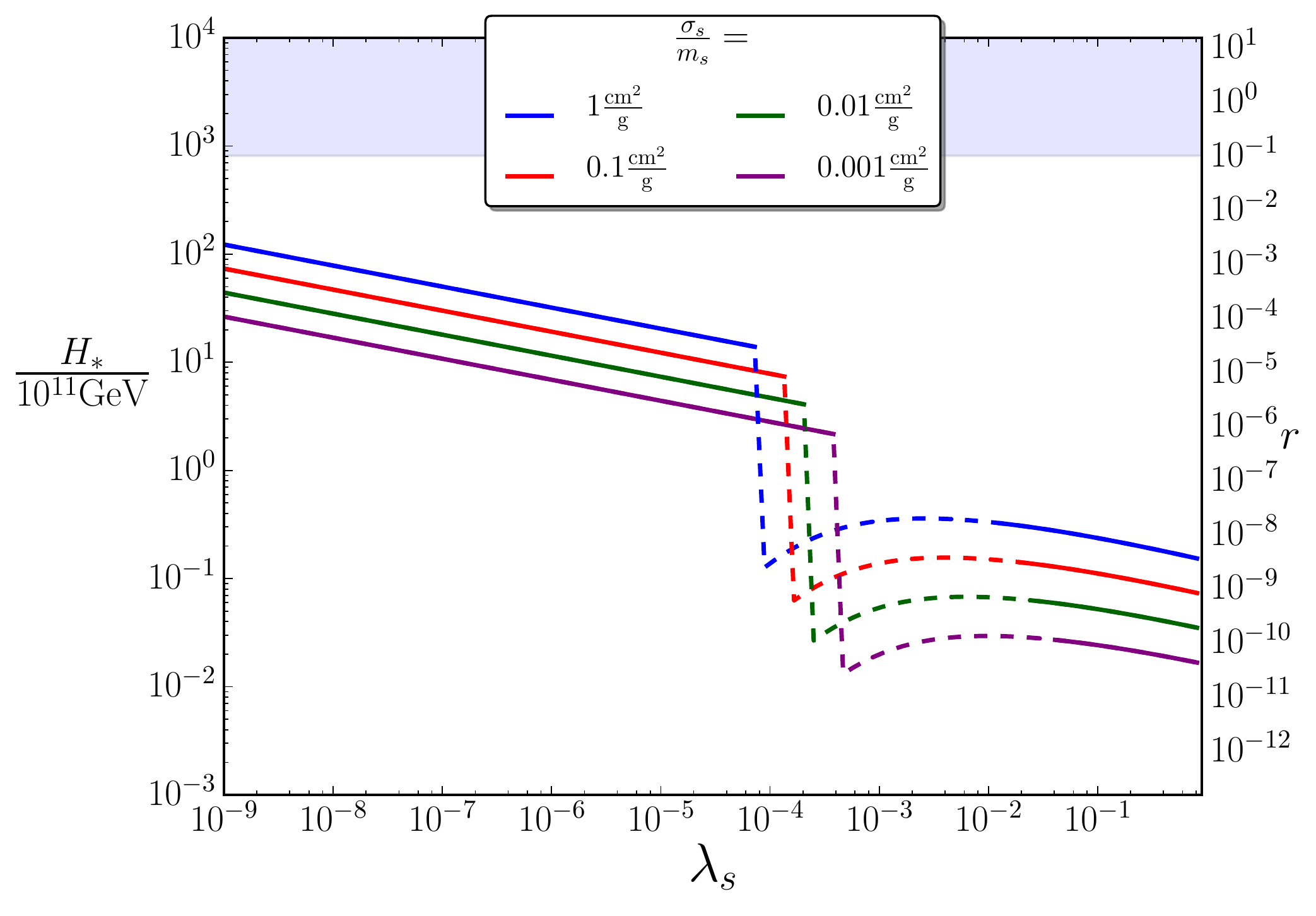}\\
\hspace{-0.7cm}\includegraphics[width=0.69\textwidth]{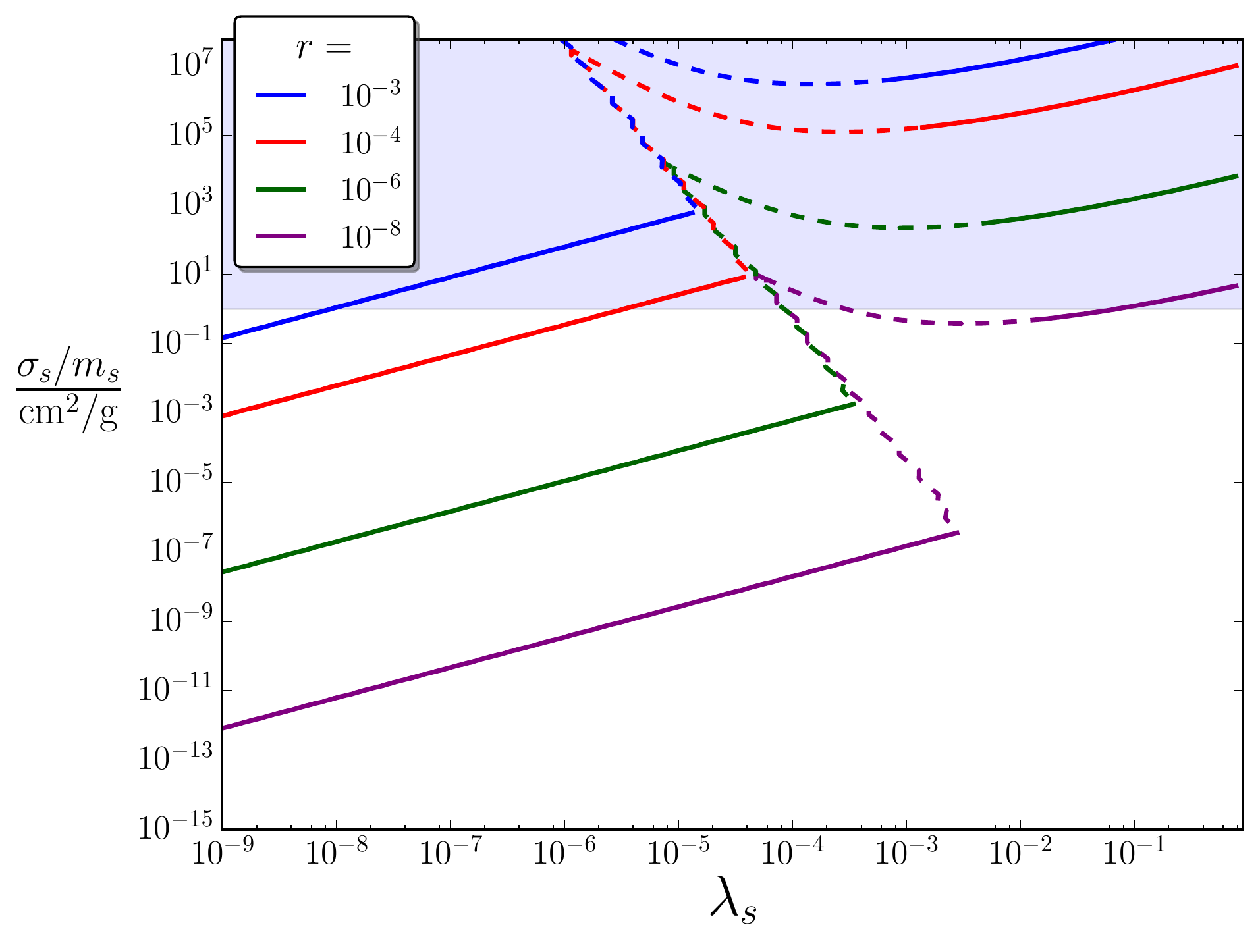}
\caption{\label{fig:constraint} 
Constraints on the tensor-to-scalar ratio $r$ for a few fixed values of the dark matter self-interaction cross-section divided by its mass, $\sigma_s/m_s$ (upper panel), and conversely, constraints on $\sigma_s/m_s$ for a few fixed values of $r$ (lower panel), as a function of the DM self-coupling constant $\lambda_s$. The dashed parts of the curves stand for the relativistic freeze-out regime (labeled ``relativistic FO'' in \Fig{fig:H_*results_mu_low}) where the calculation presented in this work does not apply. The shaded regions are observationally excluded, and correspond to the upper bound on $r$ obtained from CMB temperature and polarisation measurements~\cite{Ade:2015tva} on the upper panel and to the upper bound on $\sigma_s/m_s$ obtained from the ``Bullet Cluster'' constraint~\cite{Kaplinghat:2015aga} on the lower panel. Both plots assume that the background dynamics during inflation and reheating is standard, namely $\mu_{\rm inf}=\mu_{\rm reh}=1$, and set the DM isocurvature relative amplitude to the \emph{Planck}~\cite{Ade:2015lrj} upper limit $\beta = 0.05$ for demonstration.}
\end{center}
\end{figure}
We have presented a novel way to determine the energy scale of inflation in the case where the DM component is a feebly-interacting singlet scalar. Assuming it is light and energetically subdominant during inflation, we have shown that the inflationary energy scale $H_*$ can be expressed as a function of the DM isocurvature perturbation amplitude $\beta$ and the DM self-interaction cross-section divided by its mass $\sigma_s/m_s$, with only a very weak dependence on the DM four-point self-coupling $\lambda_s$,
\bea
\frac{H_*}{10^{11}\mathrm{GeV}}\simeq 10.0 \left(\frac{\beta}{1-\beta}\right)^{\frac{1}{3}}
\lambda_s^{-\frac{7}{36}}
\left(\frac{\sigma_s/m_s}{\mathrm{cm}^2/\mathrm{g}}\right)^{\frac{2}{9}}\, .
\eea
This relation is obtained combining \Eqs{H*_no_thermalisation} and~(\ref{scrosssection}), and is valid for the case of freeze-in only. It connects observables that constrain two seemingly unrelated topics, namely the one of inflation and the one of DM. By doing so, it opens up the possibility to access the energy scale of inflation by studying the properties of DM, and vice versa.

To illustrate this, in the upper panel of \Fig{fig:constraint} we have displayed the value of $H_*$ (and the corresponding value of the tensor-to-scalar ratio $r$) one would infer from measuring $\sigma_s/m_s$ to certain fixed values, as a function of $\lambda_s$. One can see that because of the weak dependence on $\lambda_s$, if $\sigma_s/m_s$ were measured, the energy scale of inflation would be given up to a few orders of magnitude at most, a huge improvement compared to the $15$ orders of magnitude that are a priori allowed. One should also note that a detection of $\sigma_s/m_s$ close to the current threshold~(\ref{scrosssection}) would allow one to probe values of $r$ between $10^{-9}$ and $10^{-4}$, which cannot be reached by present day CMB technology. On the lower panel conversely, we have displayed the value of the $\sigma_s/m_s$ one would infer from measuring $r$ to certain fixed values. One can see that current constraints on $\sigma_s/m_s$ already almost rule out the target of the next generation of CMB experiments $r\sim 10^{-3}$~\cite{Matsumura:2013aja, Finelli:2016cyd, DiValentino:2016foa}. In fact, if such a value were detected, then in this model $\sigma_s/m_s$ would be predicted to be close to $0.1\, \mathrm{cm}^2/\mathrm{g}$. Since this value is within the reach of forthcoming observations~\cite{Tulin:2017ara}, that would open up the possibility to either confirm or rule out the scenario presented in this work. 

In addition to presenting the basic scenario, we have also discussed the robustness of this result and quantified how it changes under various effects related to inflation, reheating, and DM dynamics at low energies. We have characterised these effects by correction factors introduced in \Eq{H*_with_corrections}. We found that the change in the background evolution during inflation and the possible thermalisation of scalar particles within the singlet sector and the following ``DM cannibalism'' phase introduce only at most $\mathcal{O}(0.1)$ and $\mathcal{O}(10^{-3})$ corrections, respectively, to the result for $H_*$, whereas variations in the reheating history can in principle have a larger effect, depending on the duration of reheating.

Although the result obtained in this work is model dependent, it is generic to a large class of scenarios and allows one to measure or constrain the energy scale of inflation even in models where the associated predicted value for the tensor-to-scalar ratio is well below the current lower bound or sensitivity of the next-generation of CMB experiments. Conversely, a detection of the tensor-to-scalar ratio would allow one to infer a measurement for the DM self-interaction cross-section. This could represent a new promising chapter in constraining DM.

\section*{Acknowledgements}
We thank M. Heikinheimo, K. Tuominen and V. Vaskonen for discussions. RJH, TT, VV and DW are supported by the UK Science and Technology Facilities Council grants ST/J001546/1, ST/N504245/1 and ST/N000668/1. VV acknowledges funding from the European Union's Horizon 2020 research and innovation programme under the Marie Sk\l odowska-Curie grant agreement N${}^0$ 750491.

\appendix

\section{Calculation of the dark matter abundance}
\label{app:general-scaling-energy-density}

In this appendix we track the energy density contained in the $s$ field from the end of inflation up until the measured abundance of dark matter today, in the sequence of events depicted in \Fig{fig:reheating}. This figure also a reference guide to the various subscripts used throughout this section.

At the end of inflation, we assume that $s_\uend$ takes a specific realisation resulting from its stochastic dynamics during inflation, $s_\uend \sim \sqrt{\langle s_\uend^2 \rangle}$, where $\langle s_\uend^2 \rangle$ is given in \Eq{eq:corrected-variance}. After the end of inflation, $s$ continues to be slowly-rolling (while quantum diffusion is shut off) until it becomes effectively massive, at the time $N_{\mathrm{osc}}$, when it starts to oscillate. One can check that the value of $s$ barely changes during this phase and to the approximation level at which the calculation is performed, it can be taken as effectively frozen, $\left.s_0\right\vert_{\mathrm{osc}}\simeq s_\uend$. The oscillations start when the effective mass of the condensate, $m_\ueff \sim \sqrt{3 \lambda_s} s_0$, becomes of order $H$. The time at which this happens can be calculated by introducing the mean equation-of-state parameter between the end of inflation and the beginning of the $s$ oscillations
\bea
\bar{w}_{\mathrm{frozen}} \equiv \frac{1}{N_{\mathrm{osc}}-N_\uend}\int^{N_{\mathrm{osc}}}_{N_\uend}w(N)\dd N\, .
\eea
The relation $\dd H/ H =-3/2(1+w) \dd N$ can then be integrated as
\bea
H_{\mathrm{osc}} = H_{\rm end} \exp\left[-\frac{3}{2}\left(1+ \bar{w}_{\mathrm{frozen}}\right) \left(N_{\mathrm{osc}}-N_\uend\right)\right]\, .
\label{eq:H:s_frozen}
\eea
By equating $H_{\rm osc} = \sqrt{3\lambda_s} \left.s_0\right\vert_{\mathrm{osc}} = \sqrt{3\lambda_s} s_\uend$, one obtains
\bea
N_{\mathrm{osc}}-N_\uend = \frac{2}{3\left(1+\bar{w}_{\mathrm{frozen}}\right)}\ln\left(\frac{H_\uend}{\sqrt{3\lambda_s}  s_\uend}\right)\, .
\label{eq:Nosc:s_frozen}
\eea
Let us note that for this number to be positive, the condition $H_\uend^2 > 3\lambda_s  s_\uend^2$ must be satisfied, which is always the case for the typical value of $s_\uend$ given by \Eq{eq:corrected-variance} if $\lambda_s\ll 1$.

After the condensate $s_0$ becomes effectively massive, it oscillates about the minimum of its quartic potential, so its energy density decays as the one of radiation, $\rho_{s_0}\propto a^{-4}$, until it fragments into $s$ particles. Fragmentation occurs when the fragmentation rate $\Gamma^{(4)}_{s_0 \rightarrow ss}$ is of order $H$. In \Ref{Tenkanen:2016idg}, it was found that
\bea
\label{gamma}
\Gamma_{s_0\rightarrow ss}^{(4)}(t)=\alpha \lambda_s^{\frac32}\sigma_0(t)\, ,
\eea
where $\alpha = 0.023$ is a numerical constant, and $\sigma_0$ is the envelope of the background $s_0$ time evolution, i.e. $s_0(t)=\sigma_0(t)\times F(t)$, where $F(t)$ is an oscillatory function. Notice that this expression is valid if the background is radiation-dominated, so reheating must have occurred at this stage for consistency. Since $\rho_{s_0}\propto s_0^4\propto a^{-4}$, during this epoch $\sigma_0 \propto 1/a$ and one has
\bea
\Gamma_{s_0\rightarrow ss}^{(4)}(N)=\alpha \lambda_s^{\frac32}   s_\uend \, \ee^{-\left(N-N_{\mathrm{osc}}\right)}\, .
\eea
On the other hand, similarly to \Eq{eq:H:s_frozen}, one has
\bea
H_{\mathrm{frag}} = H_{\mathrm{osc}} \exp\left[-\frac{3}{2}\left(1+ \bar{w}_{\mathrm{osc}}\right) \left(N_{\mathrm{frag}}-N_{\mathrm{osc}}\right)\right]\, ,
\label{eq:H:s_osc}
\eea
where $\bar{w}_{\mathrm{osc}}$ is the mean equation-of-state parameter in the oscillation phase. By equating the two previous formulas, one finds that
\begin{align}
N_{\mathrm{frag}}-N_{\mathrm{osc}} &= \frac{2}{1+3\bar{w}_{\mathrm{osc}}}\ln\left(\frac{H_{\mathrm{osc}}}{\alpha \lambda_s^{\frac32} s_\uend}\right) = -\frac{2}{1+3\bar{w}_{\mathrm{osc}}}\ln\left( \frac{{\alpha \lambda_s}}{\sqrt{3}}\right)\, ,
\label{eq:osc:Nfrag}
\end{align}
where in the second equality we have used that $H_{\rm osc} =  \sqrt{3\lambda_s} s_\uend$. One can see that in order for $N_{\mathrm{frag}}-N_{\mathrm{osc}}$ to be positive, one must have $\alpha \lambda_s < \sqrt{3}$, which is again always satisfied if $\lambda_s\ll 1$. Combining $H_{\rm osc} = \sqrt{3\lambda_s} s_\uend$, \Eqs{eq:H:s_osc} and~(\ref{eq:osc:Nfrag}), one then obtains
\bea
H_{\mathrm{frag}} = \sqrt{3\lambda_s} s_\uend \left( \frac{\alpha\lambda_s}{\sqrt{3}}\right)^{3\frac{1+\bar{w}_{\mathrm{osc}}}{1+3\bar{w}_{\mathrm{osc}}}}\, .
\eea
On the other hand, combining \Eq{eq:osc:Nfrag} with the formula $\left.\rho_{s_0}\right\vert_{\mathrm{frag}} = \rho_{s_0}\vert_{\mathrm{osc}} \ee^{-4(N_{\mathrm{frag}}-N_{\mathrm{osc}})}\simeq \left.\rho_{s_0}\right\vert_\uend \ee^{-4(N_{\mathrm{frag}}-N_{\mathrm{osc}})}$, one can further obtain
\bea
\rho_{s_0}\vert_{\mathrm{frag}} = \frac{\lambda_s}{4}  s_\uend^4 \left(\frac{\alpha \lambda_s}{\sqrt{3}}\right)^{\frac{8}{1+3\bar{w}_{\mathrm{osc}}}}\, .
\label{eq:rho_s0_frag}
\eea
Finally, let us note that at the time of fragmentation, we have assumed the singlet scalar potential to be still approximated as quartic. This means that $3\lambda_s \left.s_{0}^2\right\vert_{\mathrm{frag}} \gg m_s^2$, \ie $\rho_{s_0}\vert_{\mathrm{frag}}\gg (m_s^2/6) \sqrt{\left( \rho_{s_0}\vert_{\mathrm{frag}}\right) /\lambda_s}$, which implies the following consistency relation 
\bea \label{eq:quartic_frag_consist}
m_s \ll \sqrt{3\lambda_s}  s_\uend \left(\frac{\alpha\lambda_s}{\sqrt{3}}\right)^{\frac{2}{1+3\bar{w}_{\mathrm{osc}}}} = \left[ \frac{27\lambda_s}{2\pi^2}\frac{\Gamma^2\left( \frac{3}{4}\right)}{\Gamma^2\left( \frac{1}{4}\right)} \right]^{\frac{1}{4}}\left( \frac{\alpha \lambda_s}{\sqrt{3}}\right)^{\frac{2}{1+3\bar{w}_{\rm osc}}}H_\uend \, \mu_{\rm inf}^{-3} \, .
\eea
In the second equality, we have used \Eq{eq:corrected-variance}. As we will see below, this condition is in fact always satisfied if another condition, derived in \Eq{eq:consistency-frag-before-nrel}, is verified.

Moving on to the fragmentation products, the $s$ particles are created with a typical 3-momentum $p_s\simeq\sqrt{3\lambda_s} \left. s_0\right\vert_{\mathrm{frag}}$ \cite{Kainulainen:2016vzv}, which redshifts as the inverse of the scale factor, so that
\begin{align}
p_s &= \sqrt{3\lambda_s} \left(\frac{4\rho_{s_0}\vert_{\mathrm{frag}}}{\lambda_s}\right)^{\frac{1}{4}} \exp\left[-\left(N-N_{\mathrm{frag}}\right)\right] \nonumber \\
& = \sqrt{3\lambda_s} s_\uend \left( \frac{\alpha \lambda_s}{\sqrt{3}}\right)^{\frac{2}{1+3\bar{w}_{\mathrm{osc}}}} \exp\left[-\left(N-N_{\mathrm{frag}}\right)\right]\, ,
\end{align}
where in the second equality \Eq{eq:rho_s0_frag} has been used. When the energy becomes of order the mass $m_s$ of the particles, they stop being relativistic. This happens at the time $N_{\mathrm{nrel}}$ at which $E_s = \sqrt{m_s^2+p_s^2}\simeq m_s$ (or, roughly equivalently, when $p_s \simeq m_s$), which yields
\bea
N_{\rm nrel}- N_{\mathrm{frag}} \simeq \ln\left[  \frac{\sqrt{3\lambda_s} s_\uend}{m_s} \left( \frac{\alpha \lambda_s}{\sqrt{3}}\right)^{\frac{2}{1+3\bar{w}_{\mathrm{osc}}}}\right]\, .
\label{eq:Nms}
\eea
Requiring that $N_{\rm nrel}- N_{\mathrm{frag}}$ is positive, one finds another consistency relation, namely
\bea \label{eq:consistency-frag-before-nrel}
\frac{s_\uend }{m_s}> 3^{\frac{1-3\bar{w}_{\rm osc}}{2(1+3\bar{w}_{\rm osc})}}\alpha^{-\frac{2}{1+3\bar{w}_{\mathrm{osc}}}} \lambda_s^{-\frac{5+3\bar{w}_{\mathrm{osc}}}{2(1+3\bar{w}_{\mathrm{osc}})}}\, .
\eea
In practice, one can show that if this condition is satisfied, \Eq{eq:quartic_frag_consist} is always satisfied too. Hence, \Eq{eq:consistency-frag-before-nrel} guarantees that both consistency relations are verified, and corresponds to the grey region labeled ``no fragmentation'' in \Figs{fig:H_*results}, \ref{fig:H_*results_mu_inf}, \ref{fig:H_*results_mu_reh} and \ref{fig:H_*results_mu_low}.

During this epoch, the energy density of the $s$ particles decays as the one of radiation, so one has
\begin{align}
\rho_{s}\vert_{\rm nrel}  =  \rho_{s_0}\vert_{\mathrm{frag}} \exp\left[-4\left(N_{\rm nrel}-N_{\mathrm{frag}}\right)\right] = \frac{m_s^4}{36\lambda_s}\, ,
\end{align}
where in the second equality, we have combined \Eqs{eq:rho_s0_frag} and~(\ref{eq:Nms}). Let us also notice that since the Universe must have reheated before fragmentation in order for the result~(\ref{gamma}) to apply, at the fragmentation time it is radiation-dominated so one has $H_{\rm nrel} = H_{\mathrm{frag}} \exp[-2(N_{\rm nrel}-N_{\mathrm{frag}})]$, which gives rise to
\bea
H_{\rm nrel} = \frac{m_s^2}{\sqrt{3\lambda_s} s_\uend }\left( \frac{\alpha \lambda_s}{\sqrt{3}}\right)^{\frac{3\bar{w}_{\mathrm{osc}}-1}{3\bar{w}_{\mathrm{osc}}+1}} \,.
\label{eq:Hms}
\eea

Finally, when the particles are non-relativistic and their energy density decays as matter we can scale this up to the value it would take today, given by
\begin{equation}
\rho_{s}\vert_{\rm today} = \rho_{s}\vert_{\rm nrel}\exp\left[-3\left(N_{\rm today}-N_{\rm nrel}\right)\right] = \frac{m_s^4}{36\lambda_s} \left(\frac{a_{\rm nrel}}{a_{\rm today}}\right)^3 = \frac{m_s^4}{36\lambda_s} \left(\frac{\left.\tilde{\rho}_\gamma\right\vert_{\mathrm{today}}}{\rho_{\rm nrel}}\right)^\frac34 \,.
\end{equation}
In this expression, $\left.\tilde{\rho}_\gamma\right\vert_{\mathrm{today}}$ stands for the energy density of radiation today rescaled by the number of relativistic degrees of freedom, and $\rho_{\rm nrel}$ is the energy density of the Universe at the time when the $s$ particles became non-relativistic. This is because, as stated above, reheating must have occurred before fragmentation for consistency. Using the Friedmann equation, this gives rise to
\begin{equation}
\rho_{s}\vert_{\rm today} = \frac{m_s^4}{36\lambda_s} \left(\frac{\sqrt{\Omega_\gamma} H_{\rm today}}{H_{\rm nrel}}\right)^\frac32\, ,
\end{equation}
from which one obtains
\bea
\Omega^{(s_0)}_{\rm DM} = \frac{\rho_{s}\vert_{\rm today}}{\rho_{\rm today}} = \frac{\rho_{s}\vert_{\rm today}}{3\Mp^2 H_{\rm today}^2} = \frac{m_s^4 \Omega_\gamma^{\frac34}}{108\lambda_s\Mp^2 H_{\rm today}^2} \left(\frac{ H_{\rm today}}{H_{\rm nrel}}\right)^\frac32 \,.
\eea
By using \Eq{eq:Hms}, one finally has
\bea
\frac{\Omega^{(s_0)}_{\rm DM}h^2_{100}}{0.12} &=  0.642\, \Omega_\gamma^{\frac34}h_{100}^\frac32 \lambda_s^{-\frac14} \frac{m_s}{\rm GeV}\left(\frac{s_*}{10^{11}{\rm GeV}}\right)^{\frac{3}{2}}\left(\frac{\alpha \lambda_s}{\sqrt{3}}\right)^{\frac32 \frac{1-3\bar{w}_{\mathrm{osc}}}{1+3\bar{w}_{\mathrm{osc}}}}\, .
\eea
By comparing this expression with \Eq{eq:mu_reh_obtain_1}, one obtains the value for $\mu_{\mathrm{reh}}$ given in \Eq{eq:mu_reh}.

\section{Calculation of the portal coupling}
\label{relating_lambdahs_to_lm}

In all of the scenarios presented in this paper, we require that the $s$ particles fully constitute the DM. Through this constraint, we demonstrate here that the value of the portal coupling $\lambda_{hs}$ can be determined directly from the value of the self-interaction strength $\lambda_s$ and mass $m_s$ of the scalar field.

The time at which the dark freeze-out happens in the usual units of $x\equiv m_{s}/T_s$ is \cite{Carlson:1992fn}
\begin{equation}
x_{\rm DM}^{\rm (fo)} = \frac{m_{s}}{3.6 {\rm eV}\,\Omega^{\rm total}_{\rm DM}h_{100}^2}\frac{{\cal S}^{\rm hid}}{{\cal S}} = 2.3\times 10^9\left(\frac{m_{s}}{{\rm GeV}}\right)\frac{{\cal S}^{\rm hid}}{{\cal S}} \,,
\end{equation}
where one can compute the ratio between the entropy density of the hidden sector and that of the SM degrees of freedom, ${\cal S}^{\rm hid}/{\cal S}$, once the scalars have reached chemical equilibrium within the singlet sector, as
\begin{align}
\frac{{\cal S}^{\rm hid}}{{\rm S}} &= \frac{g^{\rm hid}_{*{\cal S}}}{g_{*{\cal S}}} \left( \frac{T_s}{T}\right)^3 = \frac{g^{\rm hid}_{*{\cal S}}(m_s)}{g_{*{\cal S}}(m_s)} \left[ \frac{g_*(m_s)}{g_*^{\rm hid}(m_s)}\frac{\rho_s(m_s)}{\rho (m_s)}\right]^{\frac{3}{4}} 
\simeq 5.3 \times 10^{8}\lambda_{hs}^{\frac{3}{2}} \, .
\end{align}
To derive this expression, we have used that
\begin{align} 
\frac{\rho_s(m_s)}{\rho (m_s)} = \frac{\rho_s(m_s)}{3H^2(m_s)\Mp^2} 
&\simeq \frac{m_sn_s(m_h)}{3H^{2}(m_h)\Mp^2 } \frac{a(m_s)}{a(m_h)}  \nonumber \\
&\simeq \frac{m_sn_h^{\rm eq}(m_h)}{H^{2}(m_h)\Mp^2}  \frac{\lambda_{hs}^2v^2}{32\pi m_h} \frac{g_{*S}^{\frac{1}{3}}(m_h)}{g_{*S}^{\frac{1}{3}}(m_s)}\frac{m_h}{m_s} \nonumber \\
&\simeq \frac{\ee^{-1}\Mp}{m_h^{3}\pi^3g_*^{\frac{1}{2}}(m_h)} \left( \frac{45}{ \pi }\right)^{\frac{3}{2}}\left( \frac{\lambda_{hs}^2v^2}{32\pi }\right) \frac{g_{*S}^{\frac{1}{3}}(m_h)}{g_{*S}^{\frac{1}{3}}(m_s)} \nonumber \\
&\simeq 9.4\times 10^{11}\lambda_{hs}^2 \,,
\end{align}
where $v=246$ GeV is the vacuum expectation value of the Higgs field and where we take $g_{*}^{\rm hid}=g_{*{\cal S}}^{\rm hid}=1$. Thus, the time of the dark freeze-out is
\begin{equation}
\label{xDFO}
x_{\rm DM}^{\rm (fo)} \simeq 1.2\times 10^{18}\lambda_{hs}^{\frac{3}{2}}\left(\frac{m_{s}}{{\rm GeV}}\right)\,.
\end{equation}

On the other hand, the dark freeze-out temperature can be estimated as the temperature at which the $4\to 2$ interaction rate drops below the Hubble rate \cite{Heikinheimo:2016yds}
\begin{equation}
\label{xDFO2}
x_{\rm DM}^{\rm (fo)} = \frac{1}{3}{\rm ln}\left[\frac{\xi^2\lambda_{s}^4\Mp}{6.5\times 10^3\sqrt{g_*}m_{s}\left(x_{\rm DM}^{\rm (fo)}\right)^{\frac{5}{2}}} \right] \,,
\end{equation}
where $\xi\equiv [g_*(m_{s})\rho_s(m_{s})/\rho(m_{s})]^{1/4}$. Equating \Eq{xDFO2} with \Eq{xDFO} and requiring $\Omega^{\rm total}_{\rm DM}h_{100}^2=0.12$ then yields a relation between the model parameters $\lambda_{hs}$, $\lambda_{s}$, $m_{s}$ and allows one to fix $\lambda_{hs}$ in terms of the other two parameters. The value we find is
\begin{equation}
\label{eq:lambdahs:ms:lambdas}
\lambda_{hs} \simeq 6.3 \times 10^{-13}\left( \frac{m_s}{{\rm GeV}}\right)^{-\frac{2}{3}} W_0^{\frac{2}{3}} \left[7.1\times 10^{4} \lambda_s^{\frac{24}{11}}\left( \frac{m_s}{{\rm GeV}}\right)^{-\frac{10}{11}} \right]  \,,
\end{equation}
where $W_0$ is the 0-branch of the Lambert W function. When plugging this expression into \Eq{mulow:1}, one obtains \Eq{mulow:2}.

\bibliography{inflationscale.bib}

\end{document}